\documentclass[10pt,journal]{IEEEtran}

\usepackage[utf8]{inputenc}
\usepackage{graphics,graphicx,color}

\usepackage[table]{xcolor}
\usepackage{xcolor, color, soul}
\usepackage{amsmath,amssymb,amsthm,amsfonts,bm,bbm}
\usepackage{algorithmicx,algorithm,algpseudocode}
\usepackage{multirow}
\usepackage[font=footnotesize, labelsep=colon]{caption}
\usepackage{cite}
\usepackage{url}
\usepackage[normalem]{ulem}
\usepackage{ragged2e}
\usepackage{graphicx}
\usepackage{framed}
\usepackage{multirow}
\usepackage{pifont} 
\usepackage{makecell} 
\usepackage{cite}
\usepackage[normalem]{ulem}
\usepackage{amsmath}
\usepackage{subcaption}
\usepackage{stfloats}
\usepackage{amsthm}
\usepackage{amssymb}
\usepackage{amsmath}
\usepackage{amsfonts}
\usepackage{enumerate}
\usepackage[utf8]{inputenc}
\usepackage{caption}
\usepackage{float}
\usepackage{hyperref}
\usepackage{tikz}
\usepackage{authblk}
\usetikzlibrary{spy}
\usepackage{pgfplots}
\usepackage{pgfplotstable}
\usepackage{graphicx,graphics,subcaption}
\usepackage[none]{hyphenat}

\usetikzlibrary{shapes.geometric, arrows}
\usetikzlibrary{automata, positioning, shapes.multipart,fit}
\tikzstyle{data} = [rectangle,  minimum width=1cm, minimum height=0.75cm,text centered, draw=gray, fill=gray!30]
\tikzstyle{process} = [circle, minimum size =1cm, text centered, draw=gray, fill=gray!30]
\tikzstyle{process_h} = [circle, minimum size =1cm, text centered, draw=blue, fill=blue!30]
\tikzstyle{arrow} = [thin,->,>=stealth]

\sethlcolor{lightgray}
\sethlcolor{yellow}

\usepackage{caption}
\usepackage{todonotes}
\presetkeys{todonotes}{color=blue!20}{}

\usepackage{color}
\definecolor{yellow}{rgb}{1,1,0.7}

\usepackage{footmisc}
\usepackage[normalem]{ulem}
\usepackage{comment}

\title{DRL-AdaPart: DRL-Driven Adaptive STAR-RIS Partitioning for Fair and Efficient Resource Utilization}
\author{Ashok S. Kumar, Nancy Nayak, \IEEEmembership{Member, IEEE}, Sheetal Kalyani, \IEEEmembership{Member, IEEE}, Himal A. Suraweera, \IEEEmembership{Senior Member, IEEE}, and Lajos Hanzo, \IEEEmembership{Life Fellow, IEEE}\\

\thanks{A. S. Kumar, and S. Kalyani are with the Department of Electrical Engineering, Indian Institute of Technology Madras, Chennai, India (e-mail: \{ee22d023@smail, skalyani@ee\}.iitm.ac.in).}
\thanks{Nancy Nayak is with the Department of Electrical and Electronic Engineering, Imperial College London, United Kingdom (e-mail: n.nayak@imperial.ac.uk).}
\thanks{H. A. Suraweera is with the Department of Electrical and Electronic Engineering, University of Peradeniya, Peradeniya 20400, Sri Lanka (e-mail: himal@eng.pdn.ac.lk).}
\thanks{Lajos Hanzo is with the School of Electronics and Computer Science, University of Southampton, SO17 1BJ Southampton, U.K. (e-mail: lh@ecs.soton.ac.uk).}

}

\begin{document}
\sloppy
	\maketitle
\begin{abstract}
Efficient resource utilization is proposed for simultaneously transmitting and reflecting reconfigurable intelligent surfaces (STAR-RIS) to ensure fair and high data rates. We optimize the number of STAR-RIS elements to be allocated to each user and maximize the sum of the user rates. To promote fairness, we introduce a soft fairness mechanism that guarantees a minimum STAR-RIS element allocation to every user. Subject to this requirement, the phase shifts of the STAR-RIS elements and the remaining element assignments are jointly optimized by harnessing an appropriately tailored deep reinforcement learning (DRL) algorithm. The proposed DRL method is also compared to Dinkelbach's algorithm and to a bespoke hybrid DRL approach. A deactivation incentive is incorporated into the DRL model for enhancing resource utilization by intelligently deactivating some of the STAR-RIS elements when not required. The proposed DRL method achieves fair and high data rates for both stationary and mobile users, while ensuring efficient resource utilization. Using the proposed DRL method, up to $34\%$ and $23\%$ of STAR-RIS elements can be deactivated in static and mobile scenarios, respectively, with negligible degradation in the average DL data rate.

	\end{abstract}
	\begin{IEEEkeywords}
    Simultaneously transmitting and reflecting reconfigurable intelligent surfaces, deep reinforcement learning, resource utilization, user fairness.
	\end{IEEEkeywords}
	
	\section{Introduction}

 Conventionally, reconfigurable intelligent surfaces (RISs) enhance the propagation environment by intelligently reflecting incident signals; however, they inherently restrict communication to $180^{\circ}$, leading to potential coverage dead zones \cite{8796365}. Addressing this limitation, the innovation of simultaneously transmitting and reflecting RIS (STAR-RIS) emerges as a transformative solution \cite{liu2021star}. Explicitly, a STAR-RIS can simultaneously reflect and transmit the incident signal, thus providing coverage to $360^{\circ}$, unlike traditional RIS, which can only reflect within the $180^{\circ}$ half-space \cite{mu2021simultaneously,10133841}. The STAR-RIS can be operated in one of the three practical modes, namely energy splitting (ES), mode switching (MS), and time switching (TS) \cite{mu2021simultaneously,liu2021star}. The MS is preferred in practice among these three protocols due to its ease of hardware implementation. The ES protocol involves optimizing more STAR-RIS coefficients, which increases computational complexity \cite{10284920}. In addition, the ES protocol is somewhat restricted by the transmission-reflection phase correlation constraint, leading to performance erosion \cite{mu2021simultaneously,liu2021star, 10284920}. By contrast, the MS protocol is not restricted by such correlation, and hence, in this work, we consider the MS STAR-RIS protocol.
\subsection{State of the Art}
The efficient utilization of resources in STAR-RIS is crucial in practical scenarios, since it reduces power consumption, minimizes operational costs, and ensures reliable communication \cite{10445726, wu2022resource}. Most of the existing work discusses the allocation of STAR-RIS resources in terms of power, beamforming, channel, and time allocation \cite{wu2022resource, xie2023performance, zhang2023resource}. Wu \textit{et al.}~\cite{wu2022resource} explore resource allocation of STAR-RIS in terms of channel assignment, power allocation, and beamforming to maximize sum rate for orthogonal multiple access. They also refine the decoding order, beamforming vector, and power allocation of non-orthogonal multiple access. Xie \textit{et al.}~\cite{xie2023performance} perform resource allocation by jointly optimizing time allocation and power allocation using a method based on genetic algorithms to maximize the total throughput. Zhang \textit{et al.}~\cite{zhang2023resource} aim for optimizing resource allocation in terms of time, power, and data offloading in a mobile edge computing system aided by a STAR-RIS for iteratively minimizing total energy consumption. However, to the best of our knowledge, none of the previous studies delves into the fact that each downlink user (DU) might require a different number of STAR-RIS elements in order to achieve a high and fair data rate. In this work, we determine the number of STAR-RIS elements required per user to achieve a fair and high data rate among all DUs. We refer to this quantity as the element assignment variable, which indicates how many elements should be assigned to serve each of the DUs to ensure fair data rates.

The first part of our work aims for maximizing the data rates by jointly optimizing the phase shifts of the STAR-RIS and the element assignment variable in a static DU scenario. Aldababsa \textit{et al.}~\cite{aldababsa2022star} and Zhao \textit{et al.}~\cite{zhao2022simultaneously} discuss a multi-user STAR-RIS setup, in which the elements of the STAR-RIS have been partitioned among the DUs so that the system resources can be utilized more effectively. Equal partitioning of the STAR-RIS allots an equal number of elements to all the users. However, an equal partitioning of the elements may result in resource over-utilization or under-utilization as users located closer to the STAR-RIS require fewer elements compared to those at greater distances. This observation motivates the proposal of an unequal partitioning strategy for the STAR-RIS to enhance resource utilization efficiency. By allocating more elements to distant users, their data rates can be improved, while assigning fewer elements to nearby users allows better exploitation of the limited number of STAR-RIS elements. In summary, a well-designed unequal partitioning scheme enhances the efficient utilization of elements while ensuring fair data rates among users. This work examines the impact of both equal and unequal partitioning, particularly in a multi-user scenario. The conventional approach for achieving fairness among DUs is to maximize the minimum of the rates; however the max-min optimization is difficult to solve in real-world applications since the landscape can be highly non-convex-non-concave, there can be multiple local saddle points, deceptive plateaus, and ridges where algorithms may get stuck \cite{razaviyayn2020nonconvex}. Maximizing the minimum rate also disregards the sum rate performance of the network, so that fairness is obtained at the expense of the achievable sum rate. To address these issues, we handle fairness by introducing a soft fairness mechanism into the sum-rate maximization problem via the per-DU element assignment constraints.
\begin{table*}[ht]
\caption{Comparison of our work with state-of-the-art STAR-RIS communication systems.}
\label{tab:allmethods}

\centering
\begin{tabular}{|>{\centering\arraybackslash}p{1.8cm}|
                >{\centering\arraybackslash}p{0.75cm}|
                >{\centering\arraybackslash}p{1.5cm}|
                >{\centering\arraybackslash}p{1.5cm}|
                >{\centering\arraybackslash}p{0.65cm}|
                >{\centering\arraybackslash}p{0.8cm}|
                >{\centering\arraybackslash}p{2.3cm}|
                >{\centering\arraybackslash}p{2.3cm}|
                >{\centering\arraybackslash}p{2cm}|} 
\hline
\textbf{Reference} & \textbf{BS-UE link} & \textbf{Channel (between STAR-RIS and BS)} & \textbf{Multi users} & \textbf{Static users} & \textbf{Mobile users} & \textbf{Partitioning of STAR-RIS elements} & \textbf{Exploring element assignment variable for fair data rates} & \textbf{Selective deactivation of STAR-RIS elements} \\
\hline
\cite{aldababsa2022star, vu2023star}   &  & Rayleigh & \ding{51} &  \ding{51} &    &  Equal  &  &    \\  
\hline

\raisebox{0pt}{\cite{aldababsa2023performance}} &   &  Nakagami & 2 users &  \ding{51} &    &  Equal  &  &   \\
\hline
\raisebox{0pt}{\cite{karim2023performance}} &  \ding{51} &  Nakagami & \ding{51} &  \ding{51} &    &  None  &  &   \\
\hline
\raisebox{0pt}{\cite{luo2024sum}} &   &  Rician & \ding{51} &  \ding{51} &   &  None  &  &   \\
\hline

\makecell{\cite{meng2023sum, wang2024joint}} & \ding{51} & Rician & \ding{51} & \ding{51} &  & None &  &  \\  
\hline

\raisebox{0pt}{\cite{eghbali2024beamforming}} &  \ding{51} &  Rician & \ding{51} &   &  \ding{51}  &  None  &  &   \\
\hline
\raisebox{0pt}{\cite{shang2025enhanced}} &  &  Rayleigh & 2 users &   &  \ding{51}  &  Equal  &  &   \\
\hline
\raisebox{0pt}{Our work} &  \ding{51} &  Rician & \ding{51} &  \ding{51} &  \ding{51}  &  Equal \& Unequal  & \ding{51} &  \ding{51}\\
\hline

\end{tabular}

\end{table*}

The solutions reported in \cite{xue2023max, perera2022sum} use traditional optimization algorithms to solve the non-convex optimization problem of finding the phase shifts of the RIS and beamformers at the base station (BS). The complexity of such algorithms increases in the face of unknown communication scenarios and for an increasing number of users and STAR-RIS elements \cite{xue2023max}. Deep reinforcement learning (DRL) is capable of solving such complex, high-dimensional problems by formulating them as a Markov decision process (MDP). Many authors explore the advantage of the DRL approach in RIS for predicting the phase shifts of the RIS, as exemplified by active beamformers at the BS \cite{nayak2024drl,ma2023deep,wang2024joint}, maximizing the weighted sum rate of uplink (UL) and downlink (DL) users, and maximizing coverage and capacity \cite{gao2023drl}. Given that the issue with the element assignment variable is inherently a combinatorial problem, we are inspired to exploit the power of DRL for optimizing both the phase shifts of the STAR-RIS and the element assignment variables.

The authors of \cite{9739892} and \cite{9500699} consider the scenario of mobile users. Explicitly, they propose an RIS-assisted system that handles users' mobility. In the context of mobile users, equal partitioning is unable to adapt to the diverse needs of each user. As a remedy, unequal partitioning allows dynamic allocation of the STAR-RIS elements in support of user-specific rate-requirements. Thus, unequal partitioning of elements is more effective in mobile scenarios than equal partitioning. In mobile user scenarios, traditional optimization algorithms face greater challenges in optimizing the phase shifts of the RIS and beamformers at the BS, since the channel characteristics change over time \cite{ma2023deep}. Our proposed DRL technique is eminently well-suited for both static and mobile scenarios.

We also investigated the effect of selective deactivation of the STAR-RIS elements on our system's performance. We demonstrate that by adopting the selective deactivation of STAR-RIS elements, the overall power consumption of the system can be reduced, and the system can further improve the resource utilization of STAR-RIS. Thus, the DRL algorithm helps the STAR-RIS make beneficial real-time decisions about the deactivation of elements, hence enhancing resource utilization and performance. In Table \ref{tab:allmethods}, we boldly contrast our techniques to the existing literature of STAR-RIS systems. 
\subsection{Our contributions}
The main contributions of our work are further detailed as follows:
\begin{itemize}
\item We propose a novel technique for achieving fair and high data rates for static and mobile users by efficiently utilizing the STAR-RIS elements in multi-user STAR-RIS DL scenarios. To achieve this objective, we introduce a new parameter, namely the element assignment variable, which determines how many STAR-RIS elements are allotted to a particular DU. By guaranteeing every DU a minimum number of allocated STAR-RIS elements, this variable embeds soft fairness directly into the sum-rate maximization problem. This enhances the flexibility of the STAR-RIS to provide a more balanced rate distribution among the DUs. Explicitly, DU fairness is assessed separately using Jain’s fairness index (JFI).
\item We formulate a joint optimization problem that optimizes the phase shift of the STAR-RIS and the element assignment variable, and then propose to solve it using an appropriately designed DRL technique.
\item Furthermore, we propose the option of selective deactivation of STAR-RIS elements to enhance resource utilization with negligible degradation in the average data rates of the DUs.
\item Our technique remains effective even in the presence of mobile DUs, where the channel conditions of the DUs fluctuate as they move, and the propagation scenario becomes highly dynamic.
\item The robustness of the proposed DRL-based joint optimization method is also evaluated under realistic imperfect channel state information (CSI) for both static and mobile DU scenarios. The proposed method shows stable convergence and reliable performance even in the presence of channel estimation errors.

\item We devised two different ways of comparing the proposed DRL method. Explicitly, we compared the conceived DRL method to Dinkelbach's scheme, that solves for STAR-RIS phase shifts. The DRL method achieves a $52\%$ higher average DL data rate than Dinkelbach's algorithm. We have been further motivated to propose a hybrid approach in which the STAR-RIS phase shifts are optimized using Dinkelbach's algorithm and binary element assignment variables are iteratively optimized using the DRL method. The DRL method advocated is observed to attain a $46\%$ higher average DL data rate than the hybrid DRL approach.

\end{itemize}

The rest of the paper is organized as follows. Section \ref{sec:System model} presents the system model for multi-user STAR-RIS systems. We detail the proposed technique in Sections \ref{sec:Deep Reinforcement Learning based solution} and \ref{sec:hybrid}. Section \ref{sec:Results and discussion} provides the simulation results, followed by our conclusions in Section \ref{sec:Conclusion}.

\emph{Notation:} Scalars are represented in lowercase, while vectors and matrices are denoted by lowercase bold and uppercase bold, respectively. The transpose and conjugate transpose of a vector or matrix are denoted by $(.)^T$ and $(.)^H$, respectively. Furthermore, $\mathrm{diag}$ represent the diagonal matrix and $\|.\|$ denotes the Euclidean norm. Finally, $\mathbb{C}$ and $\mathbb{R}$ represent complex and real sets of numbers, respectively.

    \section{System model}
\label{sec:System model}

\begin{figure*}[!ht]

\includegraphics[width=5.5 in]{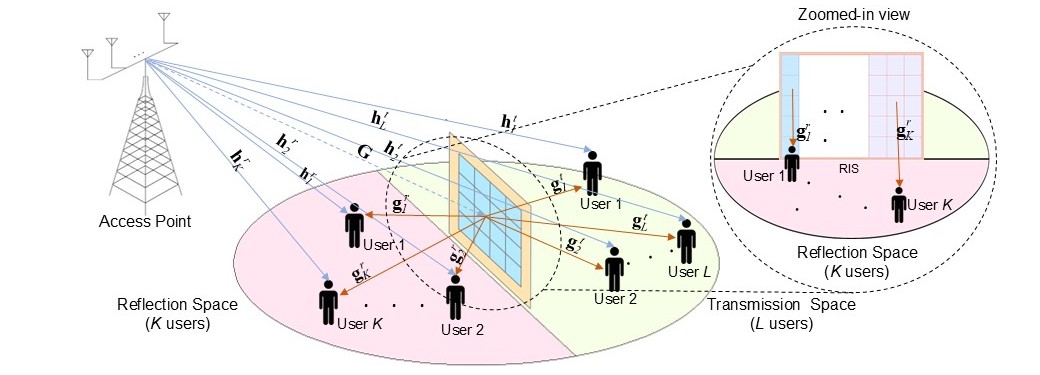}
\centering
    \caption{Proposed system model for multi-user STAR-RIS system with partitioning. We have $K$ DUs in the reflection space and $L$ DUs in the transmission space. The zoomed-in view illustrates the partitioning of the $N$ element STAR-RIS. For example, User one, being closer, is allocated fewer elements, while User $K$, being farther, is allocated more elements.}
 \label{systemmodel}   
\end{figure*}

Consider a scenario of STAR-RIS assisted DL transmission as depicted in Fig. \ref{systemmodel}. The BS has a uniform linear antenna consisting of $M$ transmit antennas. Due to blockages, direct links between the BS and DUs are assumed to be weak. Therefore, the BS communicates with the DUs with the assistance of a STAR-RIS. The STAR-RIS consists of $N=N^{h}N^{v}$ elements arranged in a uniform planar array having $N^{h}$ horizontal and $N^{v}$ vertical elements. The signal reflected from STAR-RIS remains in the same region as the incident wave (reflection space), while the transmitted signal propagates into the other region of STAR-RIS (transmission space) \cite{mu2021simultaneously}. There are $K$ DUs in the reflection space and $L$ DUs in the transmission space so that $K+L=J < N$. Let $P_{t}$ be the transmit power of the BS. Assuming equal power allocation among the $J$ DUs, the transmit power allocated to each DU is $\rho_{t} = \frac{P_{t}}{J}$ \cite{eskandari2023statistical}. Furthermore, $x_{j}$ be the data symbol intended for the $j^{th}$ DU so that $\mathbb{E}\{|x_{j}|^2\}=1$, $\forall j \in \{1,2,\dots,J\}$ \cite{meng2023sum}. The channel between the BS and STAR-RIS is given by $\mathbf{G}\in \mathbb{C}^{N \times M}$. The channel between the STAR-RIS and $k$-th ($l$-th) DU in the reflection (transmission) space is given by $\mathbf{g}_{k}^{r} \in \mathbb{C}^{1 \times N}$ $(\mathbf{g}_{l}^{t}\in \mathbb{C}^{1 \times N})$. The channel between the BS and $k$-th ($l$-th) DU in the reflection (transmission) space is given by $\mathbf{h}_{k}^{r} \in \mathbb{C}^{1\times M}$ $(\mathbf{h}_{l}^{t}\in \mathbb{C}^{1\times M})$. The reflection and transmission coefficient matrices of STAR-RIS are given by $\boldsymbol{\Theta}^r$, $\boldsymbol{\Theta}^t \in \mathbb{C}^{N\times N}$, where 
\begin{equation}
        \begin{aligned}
\boldsymbol{\Theta}^r=\mathrm{diag}\left[\sqrt{\beta_{1}^{r}} e^{j \theta_{1}^{r}}, \cdots, \sqrt{\beta_{n}^{r}} e^{j \theta_{n}^{r}}, \cdots, \sqrt{\beta_{N}^{r}} e^{j \theta_{N}^{r}}\right]^{T}
\\
\boldsymbol{\Theta}^t=\mathrm{diag}\left[\sqrt{\beta_{1}^{t}} e^{j \theta_{1}^{t}}, \cdots, \sqrt{\beta_{n}^{t}} e^{j \theta_{n}^{t}}, \cdots, \sqrt{\beta_{N}^{t}} e^{j\theta_{N}^{t}}\right]^{T}.
    \end{aligned}
\end{equation}

We consider the MS protocol for the operation of STAR-RIS, where each element is either in reflection mode or in transmission mode. For example, if the $n$-th element is allotted for a DU in the reflection space, then $\beta_{n}^{r}=1$ and $\beta_{n}^{t}=0$. Similarly, if the $n$-th element is allotted for a DU in the transmission space, then $\beta_{n}^{t}=1$ and $\beta_{n}^{r}=0$. Therefore, $\beta_{n}^{r}, \beta_{n}^{t} \in \{ 0,1 \}$, $\beta_{n}^{r}+ \beta_{n}^{t}=1$ and $\theta_{n}^{r},  \theta_{n}^{t} \in[0,2 \pi)\,\,\, \forall n\in {1,...,N}$, which characterize the amplitude and phase modifications imposed on the reflected and transmitted signal components, respectively \cite{mu2021simultaneously}.

Let $\alpha_{kn}^{r}, \alpha_{ln}^{t} \in \{0,1\}$ be the STAR-RIS element assignment variables, where $\alpha_{kn}^{r} = 1$ means that element $n$ is assigned to the $k$-th DU in the reflection space. For a specific element $n$, $\sum_{k=1}^{K} \alpha_{kn}^{r}+\sum_{l=1}^{L} \alpha_{ln}^{t} \leq 1 $, indicating that each element of the STAR-RIS should be assigned to at most one DU. However, more than one element can be assigned to one DU. Hence, the element assignment variable decides if a particular STAR-RIS element is allotted to a DU. The reflection and transmission coefficients for the $k$-th and the $l$-th DU in the reflection and transmission space after incorporating the element assignment variables are $\tilde{\boldsymbol{\Theta}}_{k}^{r}=\operatorname{diag}\left(\mathbf{\tilde{s}}_{k}^{r}\right)$, and $\tilde{\boldsymbol{\Theta}}_{l}^{t}=\operatorname{diag}\left(\mathbf{\tilde{s}}_{l}^{t}\right)$ respectively, so that
\begin{equation}
    \label{eqn:1a}
        \begin{aligned}
        \mathbf{\tilde{s}}_{k}^{r} &= \left[\alpha_{k1}^{r}\sqrt{\beta_{1}^{r}} e^{j \theta_{1}^{r}}, \cdots, \alpha_{kn}^{r}\sqrt{\beta_{n}^{r}} e^{j \theta_{n}^{r}}, \cdots, \alpha_{kN}^{r}\sqrt{\beta_{N}^{r}} e^{j \theta_{N}^{r}}\right]^{T}, \\  
    \mathbf{\tilde{s}}_{l}^{t} &= \left[\alpha_{l1}^{t}\sqrt{\beta_{1}^{t}} e^{j \theta_{1}^{t}}, \cdots,  \alpha_{ln}^{t}\sqrt{\beta_{n}^{t}} e^{j \theta_{n}^{t}}, \cdots, \alpha_{lN}^{t}\sqrt{\beta_{N}^{t}} e^{j \theta_{N}^{t}}\right]^{T}.
        \end{aligned}
\end{equation}

According to MS protocol, each element can serve a DU in either the reflection or transmission space, hence we represent the phase shift vector as
\begin{equation}
\label{eqn:2}
    \boldsymbol{\psi} = [ \theta_{1}^{\{r,t\}}, \dots,  \theta_{n}^{\{r,t\}} \dots,  \theta_{N}^{\{r,t\}}].
\end{equation} 
The superscript $\{r,t\}$ indicates that the element serves a DU in the reflection or in the transmission space. We assume that $N^{r}$ and $N^{t}$ elements are operating in the reflection and the transmission mode, respectively \cite{liu2021star} and, therefore, $N^{t}+N^{r}=N$. The element assignment matrix is $\mathbf{A}=[\mathbf{A}^{r}, \mathbf{A}^{t}]^T \in \{0,1\}^{(K+L)\times N}$, where we have
    \begin{equation}
    \label{eqn:3}
        \begin{aligned}
        \mathbf{A}^{r} &= [\boldsymbol{\alpha}^{r}_1, \boldsymbol{\alpha}^{r}_2, \dots, \boldsymbol{\alpha}^{r}_K], \,\,\mathbf{A}^{t} = [\boldsymbol{\alpha}^{t}_1, \boldsymbol{\alpha}^{t}_2, \dots, \boldsymbol{\alpha}^{t}_L] \\
        \boldsymbol{\alpha}^{r}_k = &[\alpha^{r}_{k1}, \alpha^{r}_{k2}, \dots, \alpha^{r}_{kN}]^T,\,\, 
        \boldsymbol{\alpha}^{t}_l = [\alpha^{t}_{l1}, \alpha^{t}_{l2}, \dots, \alpha^{t}_{lN}]^T
    \end{aligned}
    \end{equation}
where for a specific element $n$, $\sum_{k=1}^{K} \alpha_{kn}^{r}+\sum_{l=1}^{L} \alpha_{ln}^{t} \leq 1 $. We formulate the signal transmitted from the BS as
\begin{equation}  
\mathbf{x}=\sum_{j=1}^{J} \mathbf{w}_{j} x_j \sqrt{\rho_{t}}.
\end{equation}
Here, $\mathbf{w}_{j}  \in \mathbb{C}^{M \times 1} $ represents the beamforming vector for the $j$-th DU \cite{wang2024joint,meng2023sum}.

The total received signal at the $k$-th DU in the reflection space is

    \begin{equation}
    \begin{aligned}
         y_{k} &= (\mathbf{g}_{k}^{r} \boldsymbol{\Tilde{\Theta}}_{k}^{r}  \mathbf G + \mathbf{h}_{k}^{r} )\mathbf{w}_{k} x_k \sqrt{\rho_{t}} \\&+
     (\mathbf{g}_{k}^{r} \boldsymbol{\Tilde{\Theta}}_{k}^{r} \mathbf G + \mathbf{h}_{k}^{r} )\sum_{\substack{j=1\\ j \neq k}}^{J} \mathbf{w}_{j} x_j \sqrt{\rho_{t}} +n_{k}. 
    \end{aligned}
    \end{equation}

The signal received at the $k$-th DU comprises the desired signal, which is contributed by both the direct path from the BS and the reflection path via the STAR-RIS
\footnote{We assume that the STAR-RIS achieves directional capability to serve specific DUs via element allocation, where each group of assigned elements is coordinated to steer reflected/transmitted beams \cite{bjornson2022reconfigurable}. Under this ideal baseline assumption, unallocated elements are treated as being non-contributing in the main analysis, consistent with similar on/off models adopted in the existing RIS literature \cite{9864180, 10251590, 9605603}. In practice, unallocated elements may still introduce residual scattering. The impact of this effect on the proposed method is examined separately in Section V. G.}. It also includes interference from all other DUs in both the transmission and reflection spaces. Similarly, the total received signal at the $l$-th DU in the transmission space is

\begin{equation}
\begin{aligned}
         y_{l} &= (\mathbf{g}_{l}^{t} \boldsymbol{\Tilde{\Theta}}_{l}^{t}  \mathbf G + \mathbf{h}_{l}^{t} )\mathbf{w}_{l} x_l \sqrt{\rho_{t}} \\ &+
     (\mathbf{g}_{l}^{t} \boldsymbol{\Tilde{\Theta}}_{l}^{t} \mathbf G + \mathbf{h}_{l}^{t} )\sum_{\substack{j=1\\ j \neq l}}^{J} \mathbf{w}_{j} x_j \sqrt{\rho_{t}}+n_{l}, 
\end{aligned}
\end{equation}

\noindent where $n_{k}$ $\sim$ $\mathcal{CN}(0, \sigma_k^2)$ and $n_{l}\sim\mathcal{CN}(0, \sigma_l^2)$ \cite{mu2021simultaneously}. Here, $n_{k}$ and $n_{l}$ denote the additive white Gaussian noise at the $k$-th and $l$-th DU, respectively.

The signal-to-interference-plus-noise ratios (SINR) of the DUs denoted by $\boldsymbol{\gamma}$ are determined by the SINRs, $ \boldsymbol{\gamma}_{r}$ and $\boldsymbol{\gamma}_{t}$, corresponding to DUs in the reflection space and transmission space, respectively, where $\boldsymbol{\gamma} =\large[\boldsymbol{\gamma}_{r}, \boldsymbol{\gamma}_{t}\large]$, $\boldsymbol{\gamma}_{r} = \large[ \gamma_1^r, \gamma_2^r ,\dots , \gamma_K^r \large]$, and $\boldsymbol{\gamma}_{t} = \large[\gamma_1^t, \gamma_2^t, \dots, \gamma_L^t\large]$. The SINR $\gamma_{k}^{r}$ of DU $k$ in the reflection space  is given by
\begin{equation}     
\label{eqn:7}
    \gamma_{k}^{r}=\frac{ \rho_{t} \|\mathbf{g}_{k}^{r} \boldsymbol{\Tilde{\Theta}}_{k}^{r}  \mathbf G \mathbf{w}_{k}  + \mathbf{h}_{k}^{r} \mathbf{w}_{k}  \|^2}{ \sum\limits_{\substack{j=1\\ j \neq k}}^{J} \rho_{t} \|\mathbf{g}_{k}^{r} \boldsymbol{\Tilde{\Theta}}_{k}^{r}  \mathbf G \mathbf{w}_{j}  + \mathbf{h}_{k}^{r} \mathbf{w}_{j} \|^2+\sigma_{k}^{2}},
\end{equation}
while that of DU $l$ in the transmission space is given by
\begin{equation}
\label{eqn:8}
            \gamma_{l}^{t}=\frac{ \rho_{t} \|\mathbf{g}_{l}^{t} \boldsymbol{\Tilde{\Theta}}_{l}^{t}  \mathbf G \mathbf{w}_{l}  + \mathbf{h}_{l}^{t} \mathbf{w}_{l}   \|^2}{\sum\limits_{\substack{j=1\\ j \neq l}}^{J} \rho_{t} \|\mathbf{g}_{l}^{t} \boldsymbol{\Tilde{\Theta}}_{l}^{t}  \mathbf G \mathbf{w}_{j}+\mathbf{h}_{l}^{t} \mathbf{w}_{j}   \|^2+\sigma_{l}^{2}}.
        \end{equation}  
 
The objective is to maximize the sum of the data rates of all the DUs in both the reflection and transmission spaces, while promoting fairness by allocating more elements to distant users and fewer elements to nearby users. The DL data rate ${r}_{k}^{r}$ and ${r}_{l}^{t}$ for the $k$-th DU and for the $l$-th DU, in the reflection space and transmission space, respectively, can be expressed as
\begin{equation}
\begin{aligned}
    {r}_{k}^{r} &= \log_2(1 + \gamma_{k}^{r}), \\
    {r}_{l}^{t} &= \log_2(1 + \gamma_{l}^{t}).     
\end{aligned}
\end{equation}

\subsection{Optimization objective}

We aim to maximize the sum data rates of all DUs while inducing a fairness along the way. For this, we find the optimal element assignment and the corresponding optimal phase shift. The optimization problem is framed as follows.
\begin{subequations} \label{eq:optobjective}
    \begin{align}
    \mathcal{P}_1^{}: \quad  \underset{\boldsymbol{\psi},  \mathbf{A}}{\max} 
    & { \qquad \sum_{k=1}^{K} {r}_{k}^{r}+\sum_{l=1}^{L} {r}_{l}^{t}}  \label{eq:optobjective:obj}\\
    \text{s.t. }
    & 0 \leq \theta_{n}^{\{r,t\}} \le 2\pi \quad \forall  n\in{1,2,\dots,N} \label{eq:optimization:1a}\\
    & \sum_{n=1}^{N} \alpha_{kn}^{r} \geq 1 \quad \forall  k\in{1,2,\dots,K} \label{eq:optimization:2a}\\
    & \sum_{n=1}^{N} \alpha_{ln}^{t} \geq 1 \quad \forall l\in{1,2,\dots,L} \label{eq:optimization:3a}\\
    & \sum_{k=1}^{K} \alpha_{kn}^{r}+\sum_{l=1}^{L} \alpha_{ln}^{t} \leq 1 \,\,\,\, \forall n \in {1,\dots,N}.\label{eq:optimization:4a}
    \end{align}
\end{subequations}
Recall that $\alpha_{kn}^{r}$ is the $n$-th element assignment variable corresponding to the $k$-th DU in the reflection space, while $\alpha_{ln}^{t}$ is the $n$-th element assignment variable corresponding to $l$-th DU in the transmission space. It should be noted that $\sum_{n=1}^{N} \alpha_{kn}^{r} \geq 1$, which indicates that at least one element must be assigned to the $k$-th DU in the reflection space. Similarly, $\sum_{n=1}^{N} \alpha_{ln}^{t} \geq 1$ indicates that at least one element must be assigned to the $l$-th DU in the transmission space. The soft fairness mechanism is introduced through the constraints (\ref{eq:optimization:2a}) and (\ref{eq:optimization:3a}), which impose a minimum per-DU element assignment requirement. In the absence of these constraints, the problem becomes a sum-rate maximization problem without any minimum element assignment guarantee. In this case, assigning no elements to a DU is feasible, and the DUs rate may be significantly reduced, since it would only be served via the direct BS-DU link rather than via the STAR-RIS. Since a DU with unfavorable channel conditions contributes only a small term to the sum rate, the loss incurred by suppressing it may be outweighed by the gain obtained when its elements are reallocated to DUs having stronger channels. The sum-rate objective without the constraints (\ref{eq:optimization:2a}) and (\ref{eq:optimization:3a}), therefore, provides no incentive to serve the weakest DUs, and admits solutions in which they receive no allocation at all. In addition to that, if the contribution of the DUs with poor channels is negligible to the overall sum rate, then learning of the corresponding STAR-RIS phase shifts also becomes less important. By requiring that every DU be allocated at least one element, \eqref{eq:optimization:2a} and \eqref{eq:optimization:3a} ensure that the rate of every DU, including the weakest, contributes to the objective and remains responsive to the allocation. Fairness therefore arises from the interaction between the constraints and the objective, rather than being imposed as a separate criterion. The resulting formulation thus occupies an intermediate position between sum-rate maximization without the minimum-assignment constraints \eqref{eq:optimization:2a} and \eqref{eq:optimization:3a}, and max-min rate maximization, retaining the tractability of the former while inheriting the fairness properties of the latter.

\subsection{Design model for beamformers}
\label{beamformer}
We design the beamformers $\mathbf{w}_j,\, \forall j\in J$, using two different schemes, namely maximum ratio transmission (MRT) and zero-forcing (ZF) techniques \cite{nayak2024drl}. The MRT beamforming vectors $\mathbf{w}_{k}$ and $\mathbf{w}_{l}$ for the $k$-th and $l$-th DUs in the reflection and transmission space, respectively, are given by
\begin{equation}
\label{beamformingMRTzf}
\setlength\abovedisplayskip{2pt}
\mathbf{w}_{k}=\frac{(\mathbf{g}_{k}^{r} \boldsymbol{\Tilde{\Theta}}_{k}^{r}  \mathbf G + \mathbf{h}_{k}^{r})^{H}}{ \| \mathbf{g}_{k}^{r} \boldsymbol{\Tilde{\Theta}}_{k}^{r}  \mathbf G + \mathbf{h}_{k}^{r}\|},\,\, \mathbf{w}_{l}=\frac{(\mathbf{g}_{l}^{t} \boldsymbol{\Tilde{\Theta}}_{l}^{t}  \mathbf G + \mathbf{h}_{l}^{t})^{H}}{ \| \mathbf{g}_{l}^{t} \boldsymbol{\Tilde{\Theta}}_{l}^{t}  \mathbf G + \mathbf{h}_{l}^{t}\|}.
\end{equation}

ZF beamforming is a key technique in multi-user communication systems, specifically designed for completely eliminating inter-user interference by ensuring that the transmitted signals are orthogonal to the unintended user's channel vectors. Let \( \mathbf{W} \in \mathbb{C}^{M \times J} \) denote the ZF beamforming matrix having column vectors of \( \mathbf{w}_1, \mathbf{w}_2, \dots, \mathbf{w}_J \) and \( \mathbf{H} \in \mathbb{C}^{J \times M} \) represent the effective channel matrix. The effective channel matrix \( \mathbf{H} \in \mathbb{C}^{J \times M} \) comprises \(K\) rows corresponding to the DUs in the reflection space, each of the form \((\mathbf{g}_k^r \boldsymbol{\Tilde{\Theta}}_{k}^{r} \mathbf{G} + \mathbf{h}_k^r)\) for the $k$-th DU, and \(L\) rows corresponding to the DUs in the transmission space, each of the form \((\mathbf{g}_l^t \boldsymbol{\Tilde{\Theta}}_{l}^{t} \mathbf{G} + \mathbf{h}_l^t)\) for the $l$-th DU. The ZF beamforming matrix \( \mathbf{W} \) satisfies \( \mathbf{H} \mathbf{W} = \mathbf{I} \), where \( \mathbf{I} \) is the identity matrix \cite{madhow2008fundamentals}. The beamforming vector \( \mathbf{w}_k \) for the \(k\)-th DU in the reflection space is formulated as
\begin{equation}
    \mathbf{w}_k = \mathbf{H}^H (\mathbf{H} \mathbf{H}^H)^{-1} \mathbf{e}_k,
\end{equation}
and then normalized. Here, \( \mathbf{e}_k \) is the standard basis vector corresponding to the \(k\)-th DU. A similar expression can also be derived for DUs in the transmission space. Our objective is to focus on optimizing the STAR-RIS phase shifts and the element assignment variable for efficient resource utilization in both static and dynamic environments. The BS beamformers are determined using MRT or ZF and are not jointly optimized within the proposed DRL framework. Incorporating the BS beamforming vectors as additional optimization variables would substantially increase the DRL action space and require corresponding modifications to the state, action, and reward formulations. Therefore, the joint optimization of the BS beamformers, STAR-RIS phase shifts, and element assignments is beyond the scope of the present work and is considered as an important direction for future work.

\subsection{Channel Model}
The Rician channel models a wireless communication environment having a dominant line-of-sight (LoS) component along with multiple scattered multipath components. The channels between STAR-RIS and BS, STAR-RIS and DUs, and between BS and DUs are modeled according to Rician fading \cite{feng2020deep,chang2024star}. The channel between the BS and the $k$-th DU in the reflection space can be modeled as \cite{eghbali2024beamforming}:
\begin{equation}
\label{eqn:10}
\setlength\abovedisplayskip{12pt}
     \mathbf{h}_{k}^{r}=\sqrt{PL_{k}^{b}}\left(\sqrt{\frac{\kappa_{d}}{\kappa_{d}+1}} \mathbf{h}_{k}^{\mathrm{L}}+
    \sqrt{\frac{1}{\kappa_{d}+1}} \mathbf{h}_{k}^{\mathrm{NL}}\right). 
\setlength\belowdisplayskip{2pt}
\end{equation}

Here, $\mathbf{h}_{k}^{\mathrm{L}}$ and $\mathbf{h}_{k}^{\mathrm{NL}}$ represent the LoS and the non-LoS component of $\mathbf{h}_{k}^{\mathrm{r}}$. The LoS component $\mathbf{h}_{k}^{\mathrm{L}} \in \mathbb{C}^{1\times M}$ is given by $ \mathbf{h}_{k}^{\mathrm{L}}=e^ {\jmath 2 \pi\left(f_{k}^{b}\right) t}
\mathbf{a}_B\left(\theta_{k}^b, \phi_{k}^b\right)$,
where $f_{k}^{b}$ is the Doppler frequency shift for $\mathbf{h}_{k}^{\mathrm{r}}$. The Doppler frequency shift is given by $f_{k}^{b}=v \cos\theta_{k}^b \cos \phi_{k}^b/\lambda$, where $\lambda$ represents the wavelength. The antenna array response is given by $\mathbf{a}_B\left(\theta_{k}^b, \phi_{k}^b\right) =\left[1, \ldots, e^{-j\frac{2\pi}{\lambda}\hat{d_a}(M-1) \sin \left(\theta_{k}^b\right) \cos \left(\phi_{k}^b\right)}\right],$ where $\theta_{k}^b$ and $\phi_{k}^b$ are the azimuth and elevation angles, respectively. The distance between the antenna elements is given by $\hat{d_a}$. The non-LoS component, $\mathbf{h}_{k}^{\mathrm{NL}}\in \mathbb{C}^{1\times M}$ can be modeled as a complex Gaussian random vector $\mathcal{C N}(0,\mathbf{I}_{M})$. The distance-dependent path loss is given by $PL_{k}^{b}=C_{0}\left(d_{k}^{b}\right)^{-\zeta_{k}^{b}}$, where $C_{0}=(\frac{c}{4 \pi f_{c} d_{0}})$ denotes the path loss at the reference distance of $d_{0}= 1$m \cite{gao2023drl}. The carrier frequency is given by $f_{c}$, and $c$ represents the speed of light. Here, $d_{k}^{b}=\left\|D_{k}-D_{\mathrm{BS}}\right\|$ and $\zeta_{k}^{b} $ is the path loss exponent. Note that $D_{k}$ represents the location vector of the $k$-th DU and $D_{BS}$ denotes the location vector of the BS. In addition, $\kappa_{d}$ represents the Rician factors for $\mathbf{h}_{k}^{r}$. The channel between the STAR-RIS and $k$-th DU in the reflection space can be modeled as \cite{liu2024refracting}:
\begin{equation}
\label{eqn:11}
    \mathbf{g}_{k}^{r}=\sqrt{PL_{k}^{s}}\left(\sqrt{\frac{\xi_{d}}{\xi_{d}+1}} \mathbf{g}_{k}^{\mathrm{L}}+\sqrt{\frac{1}{\xi_{d}+1}} \mathbf{g}_{k}^{\mathrm{NL}}\right).
\end{equation}
Similarly, the channel between the BS and STAR-RIS can be formulated as
\begin{equation}
   \label{eqn:12} \mathbf{G}=\sqrt{PL_{b}^{s}}\left(\sqrt{\frac{\rho_{d}}{\rho_{d}+1}} \mathbf{G}^{\mathrm{L}}+\sqrt{\frac{1}{\rho_{d}+1}} \mathbf{G}^{\mathrm{NL}}\right).
\end{equation}
The path loss components are given by $PL_{k}^{s}=C_{0}\left(d_{k}^{s}\right)^{-\zeta_{k}^{s}}$ and $PL_{b}^{s}=C_{0}\left(d_{b}^{s}\right)^{-\zeta_{b}^{s}}$, respectively \cite{liu2024refracting}. Here $d_{k}^{s}=\left\|D_{k}-D_{S}\right\|$ and $d_{b}^{s}=\left\|D_{S}-D_{\mathrm{BS}}\right\|$. Additionally, $D_{S}$ denotes the location vector of the STAR-RIS. The LoS component $ \mathbf{g}_{k}^{\mathrm{L}} $ is given by $\mathbf{g}_{k}^{\mathrm{L}}=e^{\jmath 2 \pi\left(f_{k}^{s}\right) t}\mathbf{a}_R\left(\theta_{k}^s, \phi_{k}^s\right)$. The antenna array response vectors of STAR-RIS are given by$$
\begin{aligned}
\mathbf{a}_R\left(\theta_{k}^s, \phi_{k}^s\right)= & {\left[1, \ldots, e^{-j \frac{2\pi}{\lambda}\hat{d_s}\left(N^h-1\right) \sin \left(\theta_{k}^s\right) \cos \left(\phi_{k}^s\right)}\right] } \\
& \otimes\left[1, \ldots, e^{-j \frac{2\pi}{\lambda}\hat{d_s}\left(N^v-1\right) \sin \left(\theta_{k}^s\right) \cos \left(\phi_{k}^s\right)}\right],
\end{aligned}
$$
where $\theta_{k}^s$ and $\phi_{k}^s$ are the azimuth and elevation angles. The separation distance between the STAR-RIS elements is $\hat{d_s}$. Furthermore, $\xi_{d}$ denotes the Rician factor components of $\mathbf{g}_{k}^{r}$ and $f_{k}^{s}=v \cos\theta_{k}^s \cos \phi_{k}^s/\lambda$ represents the Doppler shift for $ \mathbf{g}_{k}^{r}$. The influence of Doppler is not considered in the channel between the BS and STAR-RIS, since both are stationary. Therefore, $\mathbf{G}^{\mathrm{L}}$ is given by $\mathbf{G}^{\mathrm{L}}=\mathbf{a}_R^H\left(\theta_{b}^s, \phi_{b}^s\right) \mathbf{a}_B\left(\theta_{s}^b, \phi_{s}^b\right)$, where $\theta_{b}^s (\theta_{s}^b)$ and $\phi_{b}^s (\phi_{s}^b)$ are the azimuth and elevation angles. Additionally, $\rho_{d}$ represents the Rician factor component of $\mathbf{G}$. Furthermore, note that $\mathbf{g}_{k}^{\mathrm{NL}} \sim \mathcal{C N}(0,\mathbf{I}_{N})$, and $\mathbf{G}^{\mathrm{NL}} \sim \mathcal{C N}(0,\mathbf{I}_{NM})$.  Similar expressions can also be derived for DUs within the transmission space. The path loss components in the channel from the BS to STAR-RIS and from STAR-RIS to the DUs are $\zeta_{b}^{s}=\zeta_{k}^{s}=2.2$ \cite{rappaport2024wireless}, while that from the BS to the DU is $\zeta_{k}^{b}=3.35$ \cite{rappaport2024wireless}.

\section{Deep Reinforcement Learning-Based Resource Utilization for Achieving Fair Data Rates}
\label{sec:Deep Reinforcement Learning based solution}
The optimization problem $\mathcal{P}_1^{}$ proposed in (\ref{eq:optobjective}) maximizes the sum of the data rates of all DUs in both the reflection and transmission spaces. Additionally, we propose a method to enhance resource utilization by selectively deactivating STAR-RIS elements. In both cases, the optimization objective is non-convex with integer constraints on the element assignment variable. A similar mixed-integer non-convex problem was addressed in \cite{perera2022sum}, where the authors have converted it to a convex problem. This was achieved by relaxing the binary constraint using a penalty-based approach and adding it to the objective function. Then it is solved using successive convex approximation, resulting in a suboptimal solution. This type of solution may not be appropriate if the channel conditions of the DUs vary a lot and the solution has to be calculated at every timestep. In contrast to this, DRL algorithms do not have to relax the integer constraints and can handle high-dimensional as well as combinatorial optimization problems more effectively \cite{zhang2023drl, aung2024aerial}. A mixed integer nonlinear programming (MINLP) problem for user scheduling, phase shift, and beamforming optimization is illustrated in \cite{huang2022joint}, where DRL is shown to outperform alternating optimization in terms of both the aggregated throughput and runtime. In our work, the non-convex integer constraint problem considered becomes non-stationary when the DUs are mobile, since it creates a dynamic environment where the positions and channel conditions of DUs perpetually change. The proposed DRL algorithm can predict at every time step, the best element assignment variables based on the channel conditions and how far the mobile DU is from STAR-RIS.

\begin{figure}
    \centering
    \includegraphics[width=3in, height=2.7in]{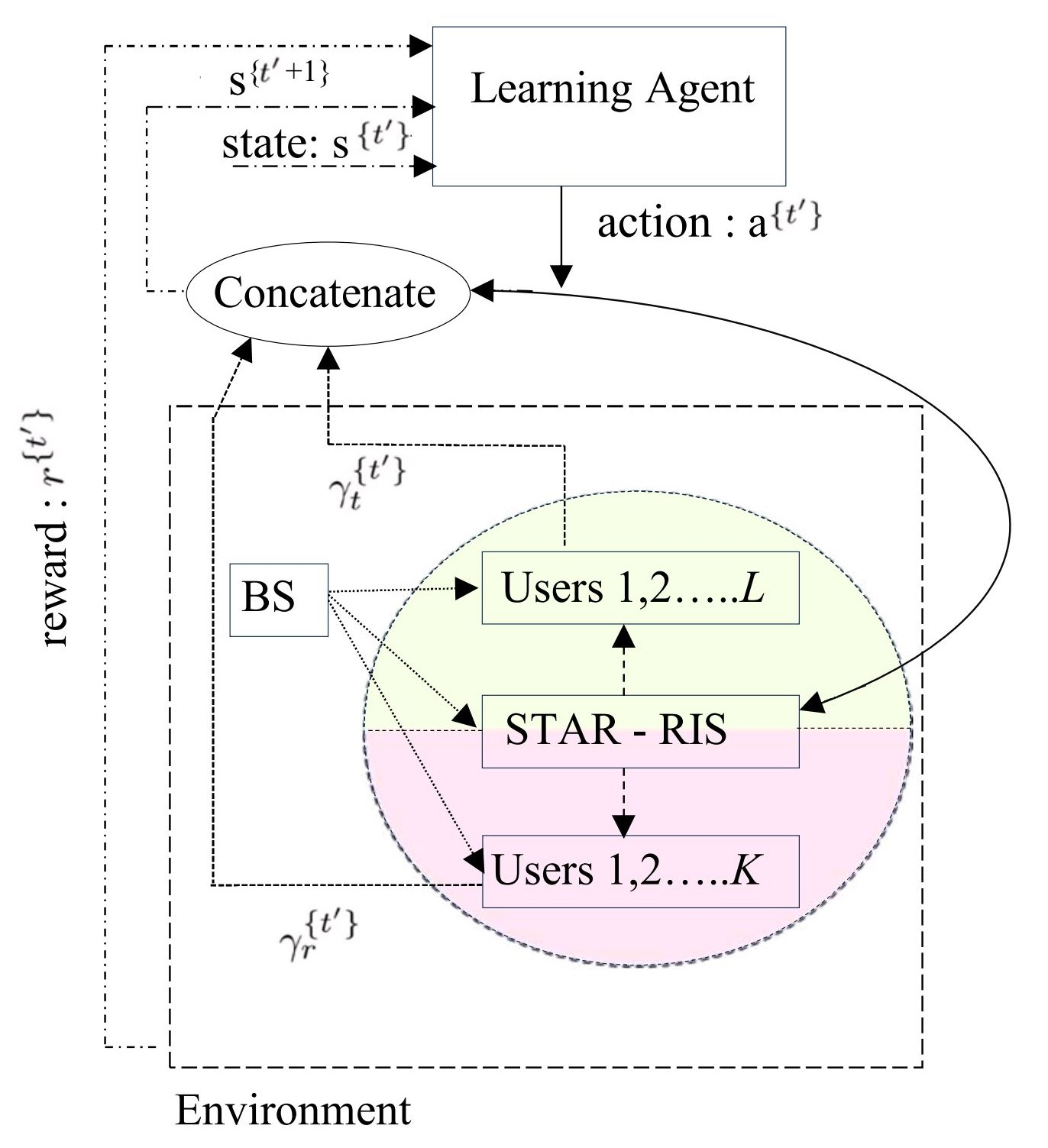}
    \caption{MDP formulation in STAR-RIS system with DRL-based learning agent.}
    \label{drl}   
\end{figure}

We model the problem of estimating the STAR-RIS phase shifts and element assignment variables as an MDP with a state space ($\mathcal{S}$), an action space ($\mathcal{A}$), an initial state distribution $\textit{p}({\mathbf{s}}^{\{ 1\}})$, a stationary state transition distribution following $\textit{p}({\mathbf{s}}^{\{ t'+1\}}|{\mathbf{s}}^{\{ t'\}}, {\mathbf{a}}^{\{ t'\}})=\textit{p}({\mathbf{s}}^{\{ t'+1\}}|{\mathbf{s}}^{\{ t'\}}, {\mathbf{a}}^{\{ t'\}}, \dots,{\mathbf{s}}^{\{ 1\}}, {\mathbf{a}}^{\{ 1\}})$ adhering to the Markov property \cite{feng2020deep,9766078}. Here, the superscript ${t'}$ indicates the values at the time step $t'$. The learning process is assessed by the reward function $r^{\{t'\}}:\mathcal{S}\times \mathcal{A} \rightarrow \mathbb{R}$, which evaluates the effectiveness of state-action pairs \cite{9766078}.  Fig. \ref{drl} shows the MDP formulation in the STAR-RIS system with a DRL-based learning agent. Here, the learning agent (deployed in the BS) takes the state ${\mathbf{s}}^{\{ t'\}}$ as its input. It then generates the action ${\mathbf{a}}^{\{ t'\}}$, which determines the phase shift of the STAR-RIS and the element assignment variable. The action predicted for the time step $t'$ can be expressed as $ \mathbf{a}^{\{t'\}}=  \left[\boldsymbol{\psi}^{\{t'\}}, \bar{\boldsymbol{\alpha}}^{\{t'\}}\right]$, while the actions $\boldsymbol{\psi}^{\{t'\}}$ are formulated using Eq. (\ref{eqn:2}). The action corresponding to the element assignment is $\bar{\boldsymbol{\alpha}}^{\{t'\}}\in \mathbb{R}^{N}$. The $n^{th}$ value of $\bar{\boldsymbol{\alpha}}^{\{t'\}}$ is mapped appropriately to represent which specific user the $n^{th}$ element of the STAR-RIS is allocated to, resulting in the element assignment matrix $\mathbf{A}$ given in Eq. (\ref{eqn:3}). The actions predicted by the model guide the transmission of signals from the BS to the DUs at the time step $t'$. The resultant SINRs at the DUs in both the reflection and transmission spaces, given by $\boldsymbol{\gamma}^{\{t'\}}$ (recall $\boldsymbol{\gamma}^{\{t'\}} = \left[\boldsymbol{\gamma}_{r}^{\{t'\}}, \boldsymbol{\gamma}_{t}^{\{t'\}}\right]$), indicate the effectiveness of the actions chosen. These SINRs become part of the observation and contribute to the state representation ${\mathbf{s}}^{\{ t'+1\}}$ for the next iteration of the learning agents where ${\mathbf{s}}^{\{ t'+1\}}=\left[\boldsymbol{\gamma}_{r}^{\{t'\}}, \boldsymbol{\gamma}_{t}^{\{t'\}}, \boldsymbol{\psi}^{\{t'\}}, \bar{\boldsymbol{\alpha}}^{\{t'\}}\right]$. The reward is calculated as the average data rate of all DUs in the reflection and transmission spaces. The data rate at time step $t'$ for DU $k$ in the reflection space is given by $r_{k}^{r{\{t'\}}} = \log_2(1 + \gamma_{k}^{r{\{t'\}}})$; and for DU $l$ in the transmission space is given by $r_{l}^{t{\{t'\}}} = \log_2(1 + \gamma_{l}^{t{\{t'\}}})$. The reward at time step $t'$ can be expressed as
\begin{equation}
\label{eq:rew}
 r^{\{t'\}} = \frac{1}{K + L} \left( \sum_{k=1}^{K} r_{k}^{r{\{t'\}}} + \sum_{l=1}^{L} r_{l}^{t{\{t'\}}} \right).
\end{equation}
To selectively deactivate excess STAR-RIS elements, we introduce a deactivation incentive into the above average-rate reward. For notational clarity, let $R^{\{t'\}}$ denote the average DL data rate across all DUs at time step $t'$, given by $R^{\{t'\}}
=
\frac{1}{K+L}
\left(
\sum_{k=1}^{K} r_{k}^{r\{t'\}}
+
\sum_{l=1}^{L} r_{l}^{t\{t'\}}
\right).$
The number of active STAR-RIS elements is defined as $N_{\mathrm{act}}^{\{t'\}}
=
\sum_{n=1}^{N}
\left(
\sum_{k=1}^{K}\alpha_{kn}^{r\{t'\}}
+
\sum_{l=1}^{L}\alpha_{ln}^{t\{t'\}}
\right)$. The revised selective deactivation reward is then formulated as
\begin{equation}
\label{eq:14}
r^{\{t'\}}
=
R^{\{t'\}}
\left[
1+
\underbrace{
\mu
\left(
1-\frac{N_{\mathrm{act}}^{\{t'\}}}{N}
\right)
}_{\text{deactivation incentive}}
\right],
\end{equation}
where, $1-N_{\mathrm{act}}^{\{t'\}}/N$ represents the normalized fraction of deactivated STAR-RIS elements, and $\mu\geq0$ is the deactivation weighting factor. The weighting factor $\mu$ controls the trade-off between data-rate and the deactivation of excess STAR-RIS elements. The goal of the RL agent is to learn a policy, denoted $\pi:\mathcal{S}\rightarrow\mathcal{A}$, to maximize the expected cumulative return, $\mathbb{E}[\mathbbm{r}^{\{t'\}}(\eta)|\pi]$, by considering the discounted rewards obtained at each time step $t'$. Here, $\eta$ denotes the discounting factor ($\eta \in [0,1]$) and $\mathbbm{r}^{\{t'\}}$  represents the return (or total discounted reward) starting from the time step $t'$. The return at time $t'$ is given by $\mathbbm{r}^{\{t'\}}(\eta)=\sum_{\hat{t}=t'}^\infty \eta^{\hat{t}-t'} r^{\{\hat{t}\}}(\mathbf{a}^{\{\hat{t}\}},\mathbf{s}^{\{\hat{t}\}})$.

The DRL agent is trained using an actor-critic method, namely, the deep deterministic policy gradient (DDPG) \cite{lillicrap2015continuous}. The actor-critic algorithm is effective for solving non-convex combinatorial optimization problems by leveraging an actor network to learn a policy that selects optimal STAR-RIS phase-shift configurations along with element assignment variables, while the critic network estimates the corresponding $Q$ function to provide policy gradient updates. This framework facilitates efficient adaptive optimization in the high-dimensional hybrid discrete-continuous action spaces of the STAR-RIS. The DDPG algorithm simultaneously learns the $Q$ function together with the policy. It uses the Bellman equation to learn the $Q$ function, which is then utilized for optimizing the policy. It contains four neural networks, namely, the actor network, critic network, target actor network, and target critic network. The actor network $\Pi$, which is parameterized by $\mathbf{\theta}^{a}$, takes the current state ${\mathbf{s}}^{\{ t'\}}$ of the environment as its input and produces the corresponding action ${\mathbf{a}}^{\{ t'\}}$ as its output. On the other hand, the critic network $\Delta$, which is parameterized by $\mathbf{\theta}^{c}$, takes both the state and the action as input and produces an output value that evaluates the effectiveness of the current policy \cite{lillicrap2015continuous}. In our case, the critic network is a fully connected network that gives a scalar output as $Q$-value. A pair of target networks was used to make learning more stable. The target actor network is parameterized by $\mathbf{\tilde{\theta}^{a}}$, and the target critic network is parameterized by $\mathbf{\tilde{\theta}^{c}}$. The environment gives the reward ${r}^{\{ t'\}}$ after each action. At each training time step $t'$, the current state $\mathbf{s}^{\{ i\}}$, the executed action $\mathbf{a}^{\{ i\}}$, the reward obtained ${r}^{\{ i\}}$, and the subsequent state $\mathbf{s}^{\{ i+1\}}$ are recorded as an experience $(\mathbf{s}^{\{ i\}}, \mathbf{a}^{\{ i\}}, {r}^{\{ i\}}, \mathbf{s}^{\{ i+1\}})$ in the buffer $B$ \cite{lillicrap2015continuous}. To train neural networks, $\tilde{N}$ samples are taken from $B$. This is used for computing the gradients. The loss function is given by the mean-squared Bellman error, as shown below
\begin{equation}
    \hat{L}=\frac{1}{\tilde{N}} \sum_{i=1}^{\tilde{N}}\left(\hat{y}^{\{ i\}}-\Delta \left((\mathbf{s}^{\{ i\}}, \mathbf{a}^{\{ i\}} \mid \mathbf{\theta}^{c}\right)\right)^2,
\end{equation}
where $\Delta (\mathbf{s}^{\{ i\}}, \mathbf{a}^{\{ i\}} \mid \mathbf{\theta}^{c})$ represents the predicted output value of the critic network and $\hat{y}^{\{ i\}}$ denotes the predicted return, which is given by
\begin{equation}
   \hat{y}^{\{ i\}}=r^{\{ i\}}+\eta\Delta \left(\mathbf{s}^{\{ i+1\}}, \Pi(\mathbf{s}^{\{ i+1\}} \mid \tilde{\theta^{a}})\mid 
 \tilde{\theta^{c}}\right).
\end{equation}
The parameters of the actor and critic networks are updated based on $\mathbf{\theta}^a \leftarrow \mathbf{\theta}^a+\eta_{a}\frac{1}{\tilde{N}} \sum_{i=1}^{\tilde{N}}\left(\nabla_{\theta^a}\Pi(s)\nabla_{a}\Delta(s,a)\mid_{a=\Pi(s)}\right)$ and $\mathbf{\theta}^c \leftarrow \mathbf{\theta}^c-\eta_{c}\nabla_{\theta^c}\hat{L}$, where $\eta_{a}, \eta_{c}\ll 1$ denotes the step size for the stochastic update. Additionally, the DDPG agents update their target actor and critic parameters using $\tilde{\theta^a} \leftarrow \tilde{\lambda}\theta^a+(1-\tilde{\lambda})\tilde{\theta^a}$ and $\tilde{\theta^c} \leftarrow \tilde{\lambda}\theta^c+(1-\tilde{\lambda})\tilde{\theta^c}$ with $\tilde{\lambda} \ll 1$. Furthermore, we add Gaussian noise to the predicted action for better exploration in the action space and, therefore, faster convergence. 
\subsection{Key innovations in the DDPG algorithm:}
The DDPG algorithm deals with continuous action and continuous state spaces. Our work addresses a pseudo-discrete action space, where the STAR-RIS phase shift is continuous, while the element assignment variable is discrete. Deep Q-Networks, which are suitable for handling discrete action spaces, cannot be applied directly to handle a mixed action space like ours \cite{mnih2013playing}. We use a two-part neural network to address the huge action space involving the STAR-RIS phase shift and the element assignment variable. The first part is constituted by a feed-forward feature extractor (FE) with $\tilde{L}$ linear layers, using ReLU as the activation function and layer normalization. This part feeds its output to the second part, comprising a pair of sub-networks (SN) that determine the phase shifts and the element assignment variable, respectively \cite{nayak2024drl}. At time step $t'$, the input to the FE is the state $\textbf{u}_{0}$=$\mathbf{s}^{\{ t'\}}$. Each layer in FE is represented by the weights $\textbf{W}_{l} \in \mathbb{R}^{d_{o}^{l}\times d_{i}^{l}} $ and biases $\textbf{b}_{l}\in \mathbb{R}^{d_{o}^{l}}$, where $l =1,2\dots \tilde{L}$, denotes the layer index and  $d_{o}^{l}$ is the output dimension, while $d_{i}^{l}$ is the input dimension. The output of FE $(\textbf{u}_{l}\in \mathbb{R}^{d_{o}^{l}})$ is given by $\textbf{u}_{l}$=ReLU$(\textbf{W}_{l}\textbf{u}_{l-1}+\textbf{b}_{l})$. The final output of the FE $\textbf{u}_{\tilde{L}}$ is then fed to the SNs. The first SN predicts the STAR-RIS phases having parameters $\textbf{W}_{\boldsymbol{\psi}} \in \mathbb{R}^{{N}\times d_{o}^{\tilde{L}}} $ and biases $\textbf{b}_{\boldsymbol{\psi}}\in \mathbb{R}^N$. We have $\mathbf{a}_{\boldsymbol{\psi}}$ = tanh$(\textbf{W}_{\boldsymbol{\psi}}\textbf{u}_{\tilde{L}}+\textbf{b}_{\boldsymbol{\psi}})$. Similarly, the second SN predicts the element assignment variable with parameters $\textbf{W}_{\bar{\boldsymbol{\alpha}}} \in \mathbb{R}^{{N}\times d_{o}^{\tilde{L}}} $ and biases $\textbf{b}_{\bar{\boldsymbol{\alpha}}}\in \mathbb{R}^N$. We have $\mathbf{a}_{\bar{\boldsymbol{\alpha}}}$ =tanh$(\textbf{W}_{\bar{\boldsymbol{\alpha}}}\textbf{u}_{\tilde{L}}+\textbf{b}_{\bar{\boldsymbol{\alpha}}})$. It should be noted that in the DRL algorithm, the initial predicted continuous actions follow a Gaussian distribution with mean zero and unit standard deviation \cite{nayak2024drl}.  This makes it a challenge for the DRL to predict actions that lack symmetry. So, it is essential to normalize the actions using the hyperbolic tangent (tanh) activation function at the final layer of each SN. This normalization ensures that the actions are symmetric and confined within the range of [-1, +1]. Later, we adjust the output according to the domain-specific knowledge of the STAR-RIS phases and element assignment variables by shifting and scaling.

The STAR-RIS phase shifts are obtained by shifting and scaling the corresponding actions to ensure that they lie within the range \([0, 2\pi]\) before using it in the environment. Therefore, we have $\boldsymbol{\psi} = 2\pi \times \left(\mathbf{a}_{\boldsymbol{\psi}} + 1\right)/2 $ radians. The element assignment variable $\bar{\boldsymbol{\alpha}}$ is a continuous vector output of the DRL policy, where each element corresponds to a specific STAR-RIS element and lies within the range $[-1,1]$. In general, for $J$ DUs, since each STAR-RIS element must be assigned to at most one DU, the range $[-1,1]$ is divided into $J$ equal intervals, where each interval corresponds to a unique DU. To form the element assignment matrix $\mathbf{A}$, a discretization process is applied to map the continuous values of $\bar{\boldsymbol{\alpha}}$ to binary assignments. Specifically, each STAR-RIS element $n$ is assigned to DU $j$ based on the interval in which its corresponding action value $\boldsymbol{\bar{{\alpha}}_{n}}$ (i.e., the $n$-th entry of $\bar{\boldsymbol{\alpha}}$) falls. This results in a binary matrix $\mathbf{A} \in \{0,1\}^{J \times N}$, where $\mathbf{A}_{j,n} = 1$ indicates that element $n$ is assigned to DU $j$, and all other entries in column $n$ are $0$ to ensure a unique assignment. Furthermore, in our work, we can selectively deactivate some of the STAR-RIS elements to improve resource utilization. To allow deactivation of the STAR-RIS elements, actions corresponding to the element assignment variables in the range $[-1,1]$ are uniformly divided into $J+1$ bins, where $J$ bins correspond to DU assignments and one bin is reserved for element deactivation. The width of each bin is $\Delta=\frac{2}{J+1}$. The corresponding bin boundaries are defined as $b_m=-1+m\Delta,\,\, m=0,1,\ldots,J+1$. The $J+1$ intervals are defined as $\mathcal{I}_0=[b_0,b_1]$ and $\mathcal{I}_m=(b_m,b_{m+1}],\,\, m=1,\ldots,J$. In general, the first $J$ intervals correspond to the $J$ DUs, according to a fixed ordering, while the last interval corresponds to element deactivation. Note that the above describes the general bin-mapping rule for an element with $J$ DUs. In the considered MS protocol, the DUs available to an element are governed by its predetermined operating mode. Accordingly, a reflection-mode element has $K+1$ bins, where the first $K$ bins correspond to the reflection-side DUs, while a transmission-mode element has $L+1$ bins, where the first $L$ bins correspond to the transmission-side DUs. The last bin in each case corresponds to element deactivation. We use the reward function given in Eq. (\ref{eq:14}) for enhancing the data rate while encouraging the deactivation of excess STAR-RIS elements.

\section{Proposed Hybrid DRL approach}
\label{sec:hybrid} 
The specific MINLP problem given in $\mathcal{P}_1$ of  (\ref{eq:optobjective}) involving the optimization of the continuous STAR-RIS phase shifts and the binary element assignment variable has not yet been addressed by any existing work regarding STAR-RIS \cite{perera2022sum, zhang2023drl, aung2024aerial, huang2022joint}. So, we have been further motivated to propose a hybrid technique, in which the STAR-RIS phase shifts are optimized using a conventional scheme such as Dinkelbach's algorithm \cite{kafizov2021wireless}, while the binary element assignment variables are optimized using the DRL method. The motivation for choosing Dinkelbach's method for optimizing the STAR-RIS phase shifts is that it solves optimization problems that are in fractional form, similar to ours. We compare the hybrid DRL to the proposed DRL method, which jointly optimizes both components. The proposed hybrid DRL is one of our key contributions,  providing an intermediate step between conventional and fully learning-based solutions.
\subsubsection{Phase shift optimization via Dinkelbach's algorithm}
We aim for finding the optimal phase shift $\boldsymbol{\psi}$ of the STAR-RIS, where the objective is to maximize the sum of the data rates of all DUs in both the reflection and transmission space. The resultant optimization problem is given by
 \begin{equation}
     \begin{aligned}
     \label{eqn2131}
\max_{\boldsymbol{\psi}} \quad & \sum_{k=1}^{K} \log_2 \left( 1 + 
\gamma_{k}^{r}
\right) + \sum_{l=1}^{L} \log_2 \left( 1 + 
\gamma_{l}^{t}
\right) \\
\text{s. t.} \quad & 0 \leq \theta_{n}^{\{r,t\}} \le 2\pi \quad \forall  n\in{1,2,\dots,N}.
\end{aligned}
 \end{equation}
In the above objective, $\max$ can be moved inside the $\log$ function, because $\log(1 + x)$ is strictly increasing and the optimization is carried out independently for each DU. The above phase optimization problem involves a fractional objective function, where the SINR for each DU is expressed as a quadratic ratio. To solve the resultant non-convex optimization problem, we express the objective as the general fractional quadratic program, of
\begin{equation}
\label{eqnm22}
    \max_{\mathbf{x} \in \mathcal{X}} \quad \frac{\mathbf{x}^T \mathbf{Q} \mathbf{x} + \mathbf{c}^T \mathbf{x} + d}{\mathbf{x}^T \mathbf{R} \mathbf{x} + \mathbf{b}^T \mathbf{x} + e},
\end{equation}
where $\mathbf{c}, \mathbf{b} \in \mathbb{R}^n$, $\mathbf{Q}, \mathbf{R} \in \mathbb{R}^{n \times n}$ are symmetric matrices, and $d, e \in \mathbb{R}$ \cite{alwazani2020intelligent}. The set $\mathcal{X} \subseteq \mathbb{R}^n$ is assumed to be convex. We then apply the matrix lifting technique for relaxing the quadratic form into a convex semidefinite program (SDP) and then apply Dinkelbach's algorithm for iteratively handling the resultant fractional structure \cite{zappone2015energy}.

Let us first consider the reflection space alone. The formulation below can also be extended to the transmission space, leading to joint optimization over both domains. The signal term in the numerator of $\gamma_{k}^{r}$ can be written as $\rho_{t} \left\| \mathbf{g}_k^r \boldsymbol{\Tilde{\Theta}}_k^r \mathbf{G} \mathbf{w}_k + \mathbf{h}_k^r \mathbf{w}_k \right\|^2 $= \scalebox{0.88}{$\rho_{t} \left( \mathbf({\tilde{s}}_{k}^{r})^H \operatorname{diag}(\mathbf{g}_k^r) \mathbf{G} \mathbf{w}_k + \mathbf{h}_k^r \mathbf{w}_k \right)
\left( \mathbf({\tilde{s}}_{k}^{r})^H \operatorname{diag}(\mathbf{g}_k^r) \mathbf{G} \mathbf{w}_k + \mathbf{h}_k^r \mathbf{w}_k \right)^H$}. Here, $\mathbf{\tilde{s}}_{k}^{r}$ is an $N$ dimensional vector that represents the STAR-RIS phase shift vector after incorporating the element assignment variables in the reflection space (recall Eq. (\ref{eqn:1a})). Note that,  $\mathbf{\tilde{s}}_{k}^{r}$ has non-zero entries only at the specific positions, where $\alpha_{kn}^{r} =1 \,\,\, \forall n \in {1,\dots,N}$. Now, the augmented phase vector can be written as 
$
\hat{\mathbf{s}}_{k}^{r} = \begin{bmatrix} \mathbf{\tilde{s}}_{k}^{r} \\ 1 \end{bmatrix} \in \mathbb{C}^{(N+1) \times 1}
$.  The lifted matrix is then $\mathbf{V}_{k} = \hat{\mathbf{s}}_{k}^{r} \left( \hat{\mathbf{s}}_{k}^{r} \right)^{H} \in \mathbb{C}^{(N+1) \times (N+1)}.$
Now, the numerator of $\gamma_k^r$ is expressed as $\rho_{t} \| \mathbf{g}_k^r \boldsymbol{\Tilde{\Theta}}_k^r \mathbf{G} \mathbf{w}_k + \mathbf{h}_k^r \mathbf{w}_k \|^2 = \rho_{t} \operatorname{Tr}(\widetilde{\mathbf{Q}}_k \mathbf{V}_{k})$, where
\begin{align*}
    \widetilde{\mathbf{Q}}_k =
    \begin{bmatrix}
    \mathbf{Q}_{k} & \mathbf{c}_k \\
    \mathbf{c}_k^H & d_k
    \end{bmatrix}.
\end{align*}
Here,  \( \mathbf{Q}_{k} = \left( \operatorname{diag}(\mathbf{g}_k^r) \mathbf{G} \mathbf{w}_k \right) \left( \operatorname{diag}(\mathbf{g}_k^r) \mathbf{G} \mathbf{w}_k \right)^H \in \mathbb{C}^{N \times N} \), 
        \( \mathbf{c}_{k} = \left( \operatorname{diag}(\mathbf{g}_k^r) \mathbf{G} \mathbf{w}_k \right) \left(\mathbf{h}_k^{r}\mathbf{w}_k \right)^{H} \in \mathbb{C}^{N \times 1} \), 
        \( d_{k} = \left(\mathbf{h}_k^{r}\mathbf{w}_k \right)\left(\mathbf{h}_k^{r}\mathbf{w}_k \right)^{H} \in \mathbb{R} \). Similarly, the denominator of $\gamma_{k}^r$ becomes $ \rho_{t} \operatorname{Tr}(\widetilde{\mathbf{R}}_k \mathbf{V}_{k})$ + $\sigma_{k}^2$, 
    where $\widetilde{\mathbf{R}}_k \in \mathbb{C}^{(N+1) \times (N+1)}$ is defined as 
    \begin{align*}
        \widetilde{\mathbf{R}}_k =
        \begin{bmatrix}
        \mathbf{R}_k & \mathbf{b}_k \\
        \mathbf{b}_k^H & e_k
        \end{bmatrix}.
    \end{align*}
Here, $\mathbf{R}_k = \sum_{j \ne k} \left( \operatorname{diag}(\mathbf{g}_k^r) \mathbf{G} \mathbf{w}_j \right) \left( \operatorname{diag}(\mathbf{g}_k^r) \mathbf{G} \mathbf{w}_j \right)^H \in \mathbb{C}^{N \times N}$, $\mathbf{b}_k = \sum_{j \ne k} \left( \operatorname{diag}(\mathbf{g}_k^r) \mathbf{G} \mathbf{w}_j \right) \left(\mathbf{h}_k^{r} \mathbf{w}_j \right)^{H} \in \mathbb{C}^{N \times 1}$, $e_k = \sum_{j \ne k} \left(\mathbf{h}_k^{r} \mathbf{w}_j \right)\left(\mathbf{h}_k^{r} \mathbf{w}_j \right)^{H} \in \mathbb{R}$. Note that \( \widetilde{\mathbf{Q}}_k, \widetilde{\mathbf{R}}_k \in \mathbb{C}^{(N+1) \times (N+1)} \) are Hermitian positive semidefinite matrices formed from the beamformers and channel matrices. By relaxing the rank constraint rank ($\mathbf{V}_{k})=1$, we obtain a convex SDP and solve it using Dinkelbach's algorithm \cite{kafizov2021wireless}. 
The maximization problem for the $k$-th reflection DU is:
\begin{equation}
    \max_{\mathbf{V}_{k} \succeq 0, \; [\mathbf{V}_{k}]_{N+1,N+1} = 1} \quad  \frac{\rho_{t} \operatorname{Tr}(\widetilde{\mathbf{Q}}_k \mathbf{V}_{k})}{\rho_{t} \operatorname{Tr}(\widetilde{\mathbf{R}}_k \mathbf{V}_{k})+\sigma_k^2}.
\end{equation}
This is a non-convex problem due to the fractional structure inside the logarithm as seen in Eq. (\ref{eqn2131}). To solve it efficiently, we apply Dinkelbach's algorithm. At each iteration $t$, we solve the following convex subproblem for the current values of $\hat\lambda_{k}^{(t)}$,
\begin{align*}
    \max_{\substack{
\mathbf{V}_{k} \succeq 0,\\
[\mathbf{V}_{k}]_{N+1,N+1} = 1
}} \,\,\,\,  \left(\rho_{t} \operatorname{Tr}(\widetilde{\mathbf{Q}}_k \mathbf{V}_{k}) - \hat\lambda_{k}^{(t)}(\rho_{t} \operatorname{Tr}(\widetilde{\mathbf{R}}_k \mathbf{V}_{k})+\sigma_k^2) \right).
\end{align*}
Each $\hat\lambda_{k}$ is updated iteratively as:
\begin{equation}
    \hat\lambda_{k}^{(t+1)} = \frac{\rho_{t} \operatorname{Tr}(\widetilde{\mathbf{Q}}_k \mathbf{V}_{k}^{(t)})}{\rho_{t} \operatorname{Tr}(\widetilde{\mathbf{R}}_k \mathbf{V}_{k}^{(t)})+\sigma_k^2}.
\end{equation}
The iterations continue until convergence, i.e., until Dinkelbach's residuals satisfy
\begin{equation}
    \left| \rho_{t} \operatorname{Tr}(\widetilde{\mathbf{Q}}_k \mathbf{V}_{k}^{(t)}) - \hat\lambda_{k}^{(t)} (\rho_{t} \operatorname{Tr}(\widetilde{\mathbf{R}}_k \mathbf{V}_{k}^{(t)})+\sigma_k^2) \right| \leq \tilde{\epsilon}, \quad \forall k
\end{equation}
where $\tilde{\epsilon}$ is a small positive constant.

\subsubsection{Element assignment variable optimization via DRL} 
Once we optimize the STAR-RIS phase shifts using Dinkelbach's algorithm, we predict the element assignment variables using a DRL agent. This leads to a two-stage hybrid optimization strategy. The hybrid optimization strategy follows the following sequence in every time step:
\begin{enumerate}
    \item The DRL agent predicts the element allocation $\bar{\boldsymbol{\alpha}}^{\{t'\}}$, based on the current state.
    \item At each time step $t'$, Dinkelbach's algorithm optimizes the phase vector $\boldsymbol{\psi}^{\{t'\}}$, using the element assignment variable as its input produced by the DRL agent.
    \item The resultant sum of the DU data rates is computed and used as a reward to update the DRL agent.
\end{enumerate}

\section{Results and discussion}
\label{sec:Results and discussion}

In this section, we investigate the performance of the proposed method and provide a performance comparison. The STAR-RIS, BS, and DUs are placed in a three-dimensional system similar to \cite{ni2022star,umer2023performance,eghbali2024beamforming}. The BS and the STAR-RIS are located at $(0, 0, 0)$, and $(48, 20, 3)$, respectively. We consider three DUs (DUs $1$, $2$, and $3$) in the transmission space and three in the reflection space (DUs $4$, $5$, and $6$), each considered to be in different locations relative to STAR-RIS to have different channel conditions. DU $2$ is closest to the STAR-RIS in transmission space, followed by DU $1$ and DU $3$. Similarly, DU $5$ is closest to STAR-RIS in the reflection space, followed by DU $4$ and DU $6$. The BS to STAR-RIS, STAR-RIS to DU, and BS to DU channels are modeled as Rician links, as mentioned in Section \ref{sec:System model}. Table \ref{tab:allmethods1} summarizes the key simulation and training parameters used in our experiments. At the beginning of every episode, i.e., at $t'=0$, the proposed DRL method randomly chooses actions, phase shifts $\boldsymbol{\psi}^{\{0\}}$ from a uniform random distribution $U(0, 2\pi)$ and $\bar{\boldsymbol{\alpha}}^{\{0\}}$ from $U(-1, 1)$.

\begin{table}[t]
\caption{Simulation and Training Parameters}
\label{tab:allmethods1}
\centering
\renewcommand{\arraystretch}{1.1}
\begin{tabular}{|p{4cm}|p{3cm}|}
\hline
\multicolumn{2}{|c|}{\textbf{System and Channel Parameters}} \\
\hline
Rician factor (BS–STAR-RIS, STAR-RIS–DU) & 10 \cite{feng2020deep,wu2021channel} \\
\hline
System bandwidth & 100 MHz \cite{taha2020deep} \\
\hline
Noise power density & $-174$ dBm/Hz \cite{peng2021multiuser} \\
\hline
Carrier frequency $f_c$ & 3.5 GHz \\ 
\hline
BS transmit power $P_{t}$ & 30 dBm \\
\hline
Evaluation metric & DL data rate (bps/Hz) \\
\hline
\multicolumn{2}{|c|}{\textbf{DRL Training Parameters}} \\
\hline
Discount factor $\eta$ & 0.6 \\
\hline
Replay buffer size & 10,000 \\
\hline
Actor learning rate & 0.0001 \\
\hline
Critic learning rate & 0.001 \\
\hline Total number of episodes & 210 \\
\hline
Time steps per episode & 1000 \\
\hline
Independent runs (averaged) & 4 \\
\hline
Hardware used & NVIDIA GeForce RTX 2080 Ti GPU \\
\hline
\end{tabular}
\label{tab:sim_params}
\end{table}

\begin{figure*}[t]
\centering
    \begin{subfigure}[t]{0.33\textwidth}
    \centering
    \pgfplotstableread[col sep = comma]{./data/IRS6BY6.csv}\datatableone
    \tikzstyle{mark_style} = [mark size={3.0}, mark repeat=20, mark phase=1]              
    \begin{tikzpicture}[thick,scale=1.7]
    \begin{axis}[
        width=3.9cm,
        height=4.3cm,
        xmin=0,
        xmax=210,
        ymin=0,
        ymax=10,
        grid=major,
        xlabel={Episode index},
        ylabel={DL rate (bps/Hz)},
       xlabel style={at={(0.50,0.03)}, yshift=5pt},
        ylabel style={at={(0.25,0.50)}, yshift=-5pt},
        label style={font=\tiny, scale=0.8},
        tick label style={font=\tiny, scale=0.7},
        legend pos=south west,
        legend cell align={left},
        legend style={at={(0.35,0)}, fill opacity=0.85, draw opacity=1.0, text opacity=1.0, font=\tiny}
        ]
         \addplot[black!50, solid, thin, mark=oplus*, mark size={2.0},every mark/.append style={solid, fill=cyan!100, opacity=1}, mark repeat=40, opacity=0.9
                ] 
            table [x=x_data, y expr=\thisrowno{1}/1000, col sep=comma]{\datatableone};

        \addplot[black!50, solid, thin, 
                mark=pentagon*, mark size={2.2}, every mark/.append style={solid, fill=blue!100, opacity=1}, mark repeat=40,  opacity=0.9
                ] 
            table [x=x_data, y expr=\thisrowno{2}/1000, col sep=comma]{\datatableone};

        \addplot[black!50, solid, thin, 
                mark=triangle*, mark size={2.0},every mark/.append style={solid, fill=violet!100, opacity=1}, mark repeat=40,  opacity=0.9
                ] 
            table [x=x_data, y expr=\thisrowno{3}/1000, col sep=comma]{\datatableone};

        \addplot[black!50, solid, thin, 
                mark=diamond*, mark size={2.0}, every mark/.append style={solid, fill=orange!100, opacity=1}, mark repeat=35,  opacity=0.9
                ] 
            table [x=x_data, y expr=\thisrowno{4}/1000, col sep=comma]{\datatableone};

        \addplot[black!50, solid, thin, 
                mark=o, mark size={2.0}, every mark/.append style={solid, fill=white!100, opacity=1}, mark repeat=30,  opacity=0.9
                ] 
            table [x=x_data, y expr=\thisrowno{5}/1000, col sep=comma]{\datatableone};

        \addplot[black!50, solid, thin, 
                mark=triangle*, mark size={2.0}, every mark/.append style={solid, fill=red!100, rotate=180, opacity=1}, mark repeat=37,  opacity=0.9
                ] 
            table [x=x_data, y expr=\thisrowno{6}/1000, col sep=comma]{\datatableone};

        \addplot[black!50, solid, thin, 
                mark=square*, mark size={2.0},every mark/.append style={solid, fill=green!100, opacity=1}, mark repeat=50,  opacity=0.9
                ] 
            table [x=x_data, y expr=\thisrowno{7}/1000, col sep=comma]{\datatableone};
    \end{axis}
\end{tikzpicture}
    \phantomcaption{} 
    \label{fig6by6}
\end{subfigure}%
\begin{subfigure}[t]{0.33\textwidth}
    \centering
    \pgfplotstableread[col sep = comma]{./data/IRS12BY12.csv}\datatableone
    \tikzstyle{mark_style} = [mark size={3.0}, mark repeat=20, mark phase=1]               
    \begin{tikzpicture}[thick,scale=1.7]
    \begin{axis}[
        width=3.9cm,
        height=4.3cm,
        xmin=0,
        xmax=210,
        ymin=0,
        ymax=20.5,
        grid=major,
        xlabel={Episode index},
        ylabel={DL rate (bps/Hz)},
        xlabel style={at={(0.50,0.03)}, yshift=5pt},
        ylabel style={at={(0.25,0.50)}, yshift=-5pt},
        label style={font=\tiny, scale=0.8},
        tick label style={font=\tiny, scale=0.7},
        legend pos=south west,
        legend cell align={left},
        legend style={at={(0.25,0)}, fill opacity=0.75, draw opacity=1.0, text opacity=1.0, font=\tiny}
        ]
         \addplot[black!50, solid, thin, mark=oplus*, mark size={2.0},every mark/.append style={solid, fill=cyan!100, opacity=1}, mark repeat=35, opacity=0.9
                ] 
            table [x=x_data, y expr=\thisrowno{1}/1000, col sep=comma]{\datatableone};

        \addplot[black!50, solid, thin, 
                mark=pentagon*, mark size={2.2}, every mark/.append style={solid, fill=blue!100, opacity=1}, mark repeat=35,  opacity=0.9
                ] 
            table [x=x_data, y expr=\thisrowno{2}/1000, col sep=comma]{\datatableone};

        \addplot[black!50, solid, thin, 
                mark=triangle*, mark size={2.0},every mark/.append style={solid, fill=violet!100, opacity=1}, mark repeat=40,  opacity=0.9
                ] 
            table [x=x_data, y expr=\thisrowno{3}/1000, col sep=comma]{\datatableone};

        \addplot[black!50, solid, thin, 
                mark=diamond*, mark size={2.0}, every mark/.append style={solid, fill=orange!100, opacity=1}, mark repeat=40,  opacity=0.9
                ] 
            table [x=x_data, y expr=\thisrowno{4}/1000, col sep=comma]{\datatableone};

        \addplot[black!50, solid, thin, 
                mark=o, mark size={2.0}, every mark/.append style={solid, fill=white!100, opacity=1}, mark repeat=30,  opacity=0.9
                ] 
            table [x=x_data, y expr=\thisrowno{5}/1000, col sep=comma]{\datatableone};

        \addplot[black!50, solid, thin, 
                mark=triangle*, mark size={2.0}, every mark/.append style={solid, fill=red!100, rotate=180, opacity=1}, mark repeat=37,  opacity=0.9
                ] 
            table [x=x_data, y expr=\thisrowno{6}/1000, col sep=comma]{\datatableone};

        \addplot[black!50, solid, thin, 
                mark=square*, mark size={2.0},every mark/.append style={solid, fill=green!100, opacity=1}, mark repeat=40,  opacity=0.9
                ] 
            table [x=x_data, y expr=\thisrowno{7}/1000, col sep=comma]{\datatableone};
    \end{axis}
\end{tikzpicture}
    \phantomcaption{} 
    \label{fig12by12}
\end{subfigure}%
\begin{subfigure}[t]{0.33\textwidth}
    \centering
    \pgfplotstableread[col sep = comma]{./data/IRS12BY12LEARNABLEA.csv}\datatableone
    \tikzstyle{mark_style} = [mark size={3.0}, mark repeat=20, mark phase=1]              
    \begin{tikzpicture}[thick,scale=1.7]
    \begin{axis}[
        width=3.9cm,
        height=4.3cm,
        xmin=0,
        xmax=210,
        ymin=0,
        ymax=20.5,
        grid=major,
        xlabel={Episode index},
        ylabel={DL rate (bps/Hz)},
       xlabel style={at={(0.50,0.03)}, yshift=5pt},
        ylabel style={at={(0.25,0.50)}, yshift=-5pt},
        label style={font=\tiny, scale=0.8},
        tick label style={font=\tiny, scale=0.7},
        legend pos=south west,
        legend cell align={left},
        legend style={at={(0.26,0)}, fill opacity=0.55, draw opacity=1.0, text opacity=1.0, font=\tiny, scale=0.8, nodes={scale=0.8}}
        ]
       
         \addplot[black!50, solid, thin, mark=oplus*, mark size={2.0},every mark/.append style={solid, fill=cyan!100, opacity=1}, mark repeat=50, opacity=0.9
                ] 
            table [x=x_data, y expr=\thisrowno{1}/1000, col sep=comma]{\datatableone};
        \addlegendentry{$r_{1}$};

        \addplot[black!50, solid, thin, 
                mark=pentagon*, mark size={2.2}, every mark/.append style={solid, fill=blue!100, opacity=1}, mark repeat=43,  opacity=0.9
                ] 
            table [x=x_data, y expr=\thisrowno{2}/1000, col sep=comma]{\datatableone};
        \addlegendentry{$r_{2}$};

        \addplot[black!50, solid, thin, 
                mark=triangle*, mark size={2.0},every mark/.append style={solid, fill=violet!100, opacity=1}, mark repeat=40,  opacity=0.9
                ] 
            table [x=x_data, y expr=\thisrowno{3}/1000, col sep=comma]{\datatableone};
        \addlegendentry{$r_{3}$};

        \addplot[black!50, solid, thin, 
                mark=diamond*, mark size={2.0}, every mark/.append style={solid, fill=orange!100, opacity=1}, mark repeat=48,  opacity=0.9
                ] 
            table [x=x_data, y expr=\thisrowno{4}/1000, col sep=comma]{\datatableone};
        \addlegendentry{$r_{4}$};

        \addplot[black!50, solid, thin, 
                mark=o, mark size={2.0}, every mark/.append style={solid, fill=white!100, opacity=1}, mark repeat=30,  opacity=0.9
                ] 
            table [x=x_data, y expr=\thisrowno{5}/1000, col sep=comma]{\datatableone};
        \addlegendentry{$r_{5}$};

        \addplot[black!50, solid, thin, 
                mark=triangle*, mark size={2.0}, every mark/.append style={solid, fill=red!100, rotate=180, opacity=1}, mark repeat=37,  opacity=0.9
                ] 
            table [x=x_data, y expr=\thisrowno{6}/1000, col sep=comma]{\datatableone};
        \addlegendentry{$r_{6}$};

        \addplot[black!50, solid, thin, 
                mark=square*, mark size={2.0},every mark/.append style={solid, fill=green!100, opacity=1}, mark repeat=50,  opacity=0.9
                ] 
            table [x=x_data, y expr=\thisrowno{7}/1000, col sep=comma]{\datatableone};
        \addlegendentry{$\frac{1}{6} \sum r_j$};

    \end{axis}
\end{tikzpicture}
    \phantomcaption{}
    \label{fig12by12-learn}
\end{subfigure}
\caption{For improved readability, the common legend for all three subfigures is shown only in Fig. 3(c): (a) $N=36$, (b) $N=144$, and (c) $N=144$ with learnable element assignment variable ($\mu=0$). The effect of increasing the number of STAR-RIS elements is shown in (a) and (b), where the elements are equally partitioned among the DUs irrespective of their location and channel conditions. Here, $r_j$ denotes the DL data rate of the $j$-th DU and ($\frac{1}{6} \sum_{j=1}^{6} r_j$) represents the average DL data rate across the six DUs. The average DL data rate improves when the STAR-RIS elements are partitioned based on the learnable element assignment variable, as shown in (c).}

\label{fig: equal partitioning}
\end{figure*}

\subsection{The impact of element assignment variable on achieving fair and high data rates among static DUs}
In this subsection, we study how the element assignment variable helps to achieve fair and high data rates among static DUs. We first look at a method, where each DU uses the same number of STAR-RIS elements, and the DRL algorithm predicts only the STAR-RIS phase shifts.
\subsubsection{Equal partitioning of STAR-RIS elements}
Consider the scenario of $N=36$ ($N^{t}=N^{r}=18$) elements, where every DU uses a fixed number of six elements. The data rates (bps/Hz) of the DUs located near (DU $2$ and DU $5$) the STAR-RIS are high, and the data rate of the DUs far (DU $3$ and DU $6$) from the STAR-RIS is low, as shown in Fig. \ref{fig6by6}. Therefore, equal partitioning of the elements leads to poor data rates for DUs located far from the STAR-RIS and is not fair.

We now examine the effect of increasing the STAR-RIS elements under equal partitioning. With an increased number of STAR-RIS elements, i.e., $N=144$ ($N^{t}=N^{r}=72$) each DU is served by $24$ elements. From Fig. \ref{fig12by12}, it can be seen that the average data rate of the system increases. The DL data rates of distant DUs also increase at a much higher rate. So, this scale-up effect indicates that increasing the number of STAR-RIS elements increases the average data rate and improves communication performance for all DUs. This also shows that the performance of distant DUs with poor channel conditions can be improved by increasing the number of STAR-RIS elements dedicated to them, which motivates us to use a varying number of STAR-RIS elements based on the channel condition of each DU.

\begin{figure*}
\centering
    \begin{subfigure}[t]{0.33\textwidth}
    \centering
    \pgfplotstableread[col sep = comma]{./data/IRS12BY12SHUTDOWN.csv}\datatableone
    \tikzstyle{mark_style} = [mark size={3.0}, mark repeat=20, mark phase=1]              
    \begin{tikzpicture}[thick,scale=1.7]
    \begin{axis}[
        width=3.9cm,
        height=4.3cm,
        xmin=0,
        xmax=210,
        ymin=0,
        ymax=20.2,
        grid=major,
        xlabel={Episode index},
        ylabel={DL rate (bps/Hz)},
        xlabel style={at={(0.50,0.03)}, yshift=5pt},
        ylabel style={at={(0.25,0.50)}, yshift=-5pt},
        label style={font=\tiny, scale=0.8},
        tick label style={font=\tiny, scale=0.7},
        legend pos=south west,
        legend cell align={left},
        legend style={at={(0.26,0)}, fill opacity=0.85, draw opacity=1.0, text opacity=1.0, font=\tiny, scale=0.8, nodes={scale=0.8}}
        ]
       
           \addplot[black!50, solid, thin, mark=oplus*, mark size={2.0},every mark/.append style={solid, fill=cyan!100, opacity=1}, mark repeat=50, opacity=0.9
                ] 
            table [x=x_data, y expr=\thisrowno{1}/1000, col sep=comma]{\datatableone};
        \addlegendentry{$r_{1}$};

        \addplot[black!50, solid, thin, 
                mark=pentagon*, mark size={2.2}, every mark/.append style={solid, fill=blue!100, opacity=1}, mark repeat=43,  opacity=0.9
                ] 
            table [x=x_data, y expr=\thisrowno{2}/1000, col sep=comma]{\datatableone};
        \addlegendentry{$r_{2}$};

        \addplot[black!50, solid, thin, 
                mark=triangle*, mark size={2.0},every mark/.append style={solid, fill=violet!100, opacity=1}, mark repeat=40,  opacity=0.9
                ] 
            table [x=x_data, y expr=\thisrowno{3}/1000, col sep=comma]{\datatableone};
        \addlegendentry{$r_{3}$};

        \addplot[black!50, solid, thin, 
                mark=diamond*, mark size={2.0}, every mark/.append style={solid, fill=orange!100, opacity=1}, mark repeat=48,  opacity=0.9
                ] 
            table [x=x_data, y expr=\thisrowno{4}/1000, col sep=comma]{\datatableone};
        \addlegendentry{$r_{4}$};

        \addplot[black!50, solid, thin, 
                mark=o, mark size={2.0}, every mark/.append style={solid, fill=white!100, opacity=1}, mark repeat=30,  opacity=0.9
                ] 
            table [x=x_data, y expr=\thisrowno{5}/1000, col sep=comma]{\datatableone};
        \addlegendentry{$r_{5}$};

        \addplot[black!50, solid, thin, 
                mark=triangle*, mark size={2.0}, every mark/.append style={solid, fill=red!100, rotate=180, opacity=1}, mark repeat=37,  opacity=0.9
                ] 
            table [x=x_data, y expr=\thisrowno{6}/1000, col sep=comma]{\datatableone};
        \addlegendentry{$r_{6}$};

        \addplot[black!50, solid, thin, 
                mark=square*, mark size={2.0},every mark/.append style={solid, fill=green!100, opacity=1}, mark repeat=50,  opacity=0.9
                ] 
            table [x=x_data, y expr=\thisrowno{7}/1000, col sep=comma]{\datatableone};
        \addlegendentry{$\frac{1}{6} \sum r_j$};

    \end{axis}
\end{tikzpicture}
    \phantomcaption{} 
    \label{IRS12BY12-SHUTDOWN}
\end{subfigure}%
\begin{subfigure}[t]{0.33\textwidth}
    \centering
    \pgfplotstableread[col sep = comma]{./data/COMPARE_LAMBDA.csv}\datatableone
    \tikzstyle{mark_style} = [mark size={3.0}, mark repeat=20, mark phase=1]             
    \begin{tikzpicture}[thick,scale=1.7]
    \begin{axis}[
        width=3.9cm,
        height=4.3cm,
        xmin=0,
        xmax=210,
        ymin=0,
        ymax=160,
        grid=major,
        xlabel={Episode index},
        ylabel={Number of active STAR-RIS elements},
       xlabel style={at={(0.50,0.03)}, yshift=5pt},
        ylabel style={at={(0.25,0.50)}, yshift=-5pt},
        label style={font=\tiny, scale=0.8},
        tick label style={font=\tiny, scale=0.7},
        legend pos=south west,
        legend cell align={left},
        legend style={at={(0.39,0)}, fill opacity=0.85, draw opacity=1.0, text opacity=1.0, font=\tiny, scale=0.75, nodes={scale=0.75}}
        ]
       \addplot[black!50, solid, thin, 
                mark=diamond*, mark size={2.0}, every mark/.append style={solid, fill=red!100, opacity=1.0}, mark repeat=40,  opacity=0.9
                ] 
            table [x=x_data, y expr=\thisrowno{1}, col sep=comma]{\datatableone};
        \addlegendentry{$\mu$=0};

         \addplot[black!50, solid, thin, mark=oplus*, mark size={2.0},every mark/.append style={solid, fill=cyan!100, opacity=1}, mark repeat=40, opacity=0.9
                ]
            table [x=x_data, y expr=\thisrowno{4}, col sep=comma]{\datatableone};
        \addlegendentry{$\mu$=0.1};

         \addplot[black!50, solid, thin, 
                mark=square*, mark size={2.0}, every mark/.append style={solid, fill=green!100, opacity=1.0}, mark repeat=40,  opacity=0.9
                ]  
            table [x=x_data, y expr=\thisrowno{2}, col sep=comma]{\datatableone};
        \addlegendentry{$\mu$=0.4};

         \addplot[black!50, solid, thin, 
                mark=pentagon*, mark size={2.2}, every mark/.append style={solid, fill=blue!100, opacity=1}, mark repeat=40,  opacity=0.9
                ]   
            table [x=x_data, y expr=\thisrowno{3}, col sep=comma]{\datatableone};
        \addlegendentry{$\mu$=0.6};

    \end{axis}
\end{tikzpicture}
    \phantomcaption{} 
    \label{comparelambda}
    \end{subfigure}
    \begin{subfigure}[t]{0.33\textwidth}
    \centering
    \pgfplotstableread[col sep = comma]{./data/DATARATE_FOR_DIFFERENT_LAMBDA.csv}\datatableone
    \tikzstyle{mark_style} = [mark size={3.0}, mark repeat=20, mark phase=1]             
    \begin{tikzpicture}[thick,scale=1.7]
    \begin{axis}[
        width=3.9cm,
        height=4.3cm,
        xmin=0,
        xmax=210,
        ymin=0,
        ymax=18,
        grid=major,
        xlabel={Episode index},
        ylabel={Average DL rate for different $\mu$(bps/Hz)},
        xlabel style={at={(0.50,0.03)}, yshift=5pt},
        ylabel style={at={(0.25,0.50)}, yshift=-5pt},
        label style={font=\tiny, scale=0.8},
        tick label style={font=\tiny, scale=0.7},
        legend pos=south west,
        legend cell align={left},
        legend style={at={(0.37,0)}, fill opacity=0.85, draw opacity=1.0, text opacity=1.0, font=\tiny, scale=0.8, nodes={scale=0.8}}
        ]
        \addplot[black!50, solid, thin, 
                mark=diamond*, mark size={2.0}, every mark/.append style={solid, fill=red!100, opacity=1.0}, mark repeat=50,  opacity=0.9
                ] 
            table [x=x_data, y expr=\thisrowno{1}/1000, col sep=comma]{\datatableone};
        \addlegendentry{$\mu$=0};
         \addplot[black!50, solid, thin, mark=oplus*, mark size={2.0},every mark/.append style={solid, fill=cyan!100, opacity=1}, mark repeat=55, opacity=0.9
                ]
            table [x=x_data, y expr=\thisrowno{2}/1000, col sep=comma]{\datatableone};
        \addlegendentry{$\mu$=0.1};

         \addplot[black!50, solid, thin, 
                mark=square*, mark size={2.0}, every mark/.append style={solid, fill=green!100, opacity=1.0}, mark repeat=60,  opacity=0.9
                ]  
            table [x=x_data, y expr=\thisrowno{3}/1000, col sep=comma]{\datatableone};
        \addlegendentry{$\mu$=0.4};

        \addplot[black!50, solid, thin, 
                mark=pentagon*, mark size={2.2}, every mark/.append style={solid, fill=blue!100, opacity=1}, mark repeat=50,  opacity=0.9
                ] 
            table [x=x_data, y expr=\thisrowno{4}/1000, col sep=comma]{\datatableone};
        \addlegendentry{$\mu$=0.6};

    \end{axis}
\end{tikzpicture}
    \phantomcaption{}
    \label{dataratefordifferentlambda}
\end{subfigure}

\caption{The effect of deactivating the elements of STAR-RIS for $N = 144$ in a static DU scenario: (a) Data rate plot for $\mu =0.4$ (better visible in colour). Here, $r_j$ denotes the DL data rate of the $j$-th DU and ($\frac{1}{6} \sum_{j=1}^{6} r_j$) represents the average DL data rate across the six DUs. (b) Number of active STAR-RIS elements for different $\mu$ values vs. episode index, and (c) Average data rate vs. episode index for different $\mu$ values. Notably, the average data rate remains nearly close for $\mu$ values of 0, 0.1, and 0.4.}
\label{fig: shutdown}
\end{figure*}

\begin{figure*}
\centering
    \begin{subfigure}[t]{0.33\textwidth}
    \centering
    \pgfplotstableread[col sep = comma]{./data/IRS12BY12MOBILE.csv}\datatableone
    \tikzstyle{mark_style} = [mark size={3.0}, mark repeat=20, mark phase=1]             
    \begin{tikzpicture}[thick,scale=1.7]
    \begin{axis}[
        width=3.9cm,
        height=4.3cm,
        xmin=0,
        xmax=210,
        ymin=0,
        ymax=11,
        grid=major,
        xlabel={Episode index},
        ylabel={DL rate (bps/Hz)},
        xlabel style={at={(0.50,0.03)}, yshift=5pt},
        ylabel style={at={(0.25,0.50)}, yshift=-5pt},
        label style={font=\tiny, scale=0.8},
        tick label style={font=\tiny, scale=0.7},
        legend pos=south west,
        legend cell align={left},
        legend style={at={(0.27,0)}, fill opacity=0.85, draw opacity=1.0, text opacity=1.0, font=\tiny, scale=0.75, nodes={scale=0.75}}
        ]
         \addplot[black!50, solid, thin, mark=oplus*, mark size={2.0},every mark/.append style={solid, fill=cyan!100, opacity=1}, mark repeat=51, opacity=0.9
                ] 
            table [x=x_data, y expr=\thisrowno{1}/1000, col sep=comma]{\datatableone};
        \addlegendentry{$r_{1}$};

        \addplot[black!50, solid, thin, 
                mark=pentagon*, mark size={2.2}, every mark/.append style={solid, fill=blue!100, opacity=1}, mark repeat=43,  opacity=0.9
                ] 
            table [x=x_data, y expr=\thisrowno{2}/1000, col sep=comma]{\datatableone};
        \addlegendentry{$r_{2}$};

        \addplot[black!50, solid, thin, 
                mark=triangle*, mark size={2.0},every mark/.append style={solid, fill=violet!100, opacity=1}, mark repeat=40,  opacity=0.9
                ] 
            table [x=x_data, y expr=\thisrowno{3}/1000, col sep=comma]{\datatableone};
        \addlegendentry{$r_{3}$};

        \addplot[black!50, solid, thin, 
                mark=diamond*, mark size={2.0}, every mark/.append style={solid, fill=orange!100, opacity=1}, mark repeat=48,  opacity=0.9
                ] 
            table [x=x_data, y expr=\thisrowno{4}/1000, col sep=comma]{\datatableone};
        \addlegendentry{$r_{4}$};

        \addplot[black!50, solid, thin, 
                mark=o, mark size={2.0}, every mark/.append style={solid, fill=white!100, opacity=1}, mark repeat=30,  opacity=0.9
                ] 
            table [x=x_data, y expr=\thisrowno{5}/1000, col sep=comma]{\datatableone};
        \addlegendentry{$r_{5}$};

        \addplot[black!50, solid, thin, 
                mark=triangle*, mark size={2.0}, every mark/.append style={solid, fill=red!100, rotate=180, opacity=1}, mark repeat=37,  opacity=0.9
                ] 
            table [x=x_data, y expr=\thisrowno{6}/1000, col sep=comma]{\datatableone};
        \addlegendentry{$r_{6}$};

        \addplot[black!50, solid, thin, 
                mark=square*, mark size={2.0},every mark/.append style={solid, fill=green!100, opacity=1}, mark repeat=50,  opacity=0.9
                ] 
            table [x=x_data, y expr=\thisrowno{7}/1000, col sep=comma]{\datatableone};
        \addlegendentry{$\frac{1}{6} \sum r_j$};
    \end{axis}
\end{tikzpicture}
    \phantomcaption{} 
    \label{IRS12BY12-mobile}
    \end{subfigure}%
    \begin{subfigure}[t]{0.33\textwidth}
    \centering
    \pgfplotstableread[col sep = comma]{./data/COMPARE_LAMBDADYNAMIC.csv}\datatableone
    \tikzstyle{mark_style} = [mark size={3.0}, mark repeat=20, mark phase=1]             
    \begin{tikzpicture}[thick,scale=1.7]
    \begin{axis}[
        width=3.9cm,
        height=4.3cm,
        xmin=0,
        xmax=210,
        ymin=0,
        ymax=160,
        grid=major,
        xlabel={Episode index},
        ylabel={Number of active STAR-RIS elements},
       xlabel style={at={(0.50,0.03)}, yshift=5pt},
        ylabel style={at={(0.25,0.50)}, yshift=-5pt},
        label style={font=\tiny, scale=0.8},
        tick label style={font=\tiny, scale=0.7},
        legend pos=south west,
        legend cell align={left},
        legend style={at={(0.39,0)}, fill opacity=0.85, draw opacity=1.0, text opacity=1.0, font=\tiny, scale=0.75, nodes={scale=0.75}}
        ]
       \addplot[black!50, solid, thin, 
                mark=diamond*, mark size={2.0}, every mark/.append style={solid, fill=red!100, opacity=1.0}, mark repeat=40,  opacity=0.9
                ] 
            table [x=x_data, y expr=\thisrowno{1}, col sep=comma]{\datatableone};
        \addlegendentry{$\mu$=0};

         \addplot[black!50, solid, thin, mark=oplus*, mark size={2.0},every mark/.append style={solid, fill=cyan!100, opacity=1}, mark repeat=40, opacity=0.9
                ]
            table [x=x_data, y expr=\thisrowno{2}, col sep=comma]{\datatableone};
        \addlegendentry{$\mu$=0.1};

         \addplot[black!50, solid, thin, 
                mark=square*, mark size={2.0}, every mark/.append style={solid, fill=green!100, opacity=1.0}, mark repeat=40,  opacity=0.9
                ]  
            table [x=x_data, y expr=\thisrowno{3}, col sep=comma]{\datatableone};
        \addlegendentry{$\mu$=0.3};

         \addplot[black!50, solid, thin, 
                mark=pentagon*, mark size={2.2}, every mark/.append style={solid, fill=blue!100, opacity=1}, mark repeat=40,  opacity=0.9
                ]   
            table [x=x_data, y expr=\thisrowno{4}, col sep=comma]{\datatableone};
        \addlegendentry{$\mu$=0.6};

    \end{axis}
\end{tikzpicture}
    \phantomcaption{} 
    \label{comparelambdadynamic}
    \end{subfigure}%
    \begin{subfigure}[t]{0.33\textwidth}
    \centering
    \pgfplotstableread[col sep = comma]{./data/DATARATE_DYNAMICCOMPARE.csv}\datatableone
    \tikzstyle{mark_style} = [mark size={3.0}, mark repeat=20, mark phase=1]             
    \begin{tikzpicture}[thick,scale=1.7]
    \begin{axis}[
        width=3.9cm,
        height=4.3cm,
        xmin=0,
        xmax=210,
        ymin=0,
        ymax=11.5,
        grid=major,
        xlabel={Episode index},
        ylabel={Average DL rate for different $\mu$(bps/Hz)},
        xlabel style={at={(0.50,0.03)}, yshift=5pt},
        ylabel style={at={(0.25,0.50)}, yshift=-5pt},
        label style={font=\tiny, scale=0.8},
        tick label style={font=\tiny, scale=0.7},
        legend pos=south west,
        legend cell align={left},
        legend style={at={(0.37,0)}, fill opacity=0.85, draw opacity=1.0, text opacity=1.0, font=\tiny, scale=0.8, nodes={scale=0.8}}
        ]
        \addplot[black!50, solid, thin, 
                mark=diamond*, mark size={2.0}, every mark/.append style={solid, fill=red!100, opacity=1.0}, mark repeat=50,  opacity=0.9
                ] 
            table [x=x_data, y expr=\thisrowno{1}/1000, col sep=comma]{\datatableone};
        \addlegendentry{$\mu$=0};
         \addplot[black!50, solid, thin, mark=oplus*, mark size={2.0},every mark/.append style={solid, fill=cyan!100, opacity=1}, mark repeat=55, opacity=0.9
                ]
            table [x=x_data, y expr=\thisrowno{2}/1000, col sep=comma]{\datatableone};
        \addlegendentry{$\mu$=0.1};

         \addplot[black!50, solid, thin, 
                mark=square*, mark size={2.0}, every mark/.append style={solid, fill=green!100, opacity=1.0}, mark repeat=60,  opacity=0.9
                ]  
            table [x=x_data, y expr=\thisrowno{3}/1000, col sep=comma]{\datatableone};
        \addlegendentry{$\mu$=0.3};

        \addplot[black!50, solid, thin, 
                mark=pentagon*, mark size={2.2}, every mark/.append style={solid, fill=blue!100, opacity=1}, mark repeat=50,  opacity=0.9
                ] 
            table [x=x_data, y expr=\thisrowno{4}/1000, col sep=comma]{\datatableone};
        \addlegendentry{$\mu$=0.6};

    \end{axis}
\end{tikzpicture}
    \phantomcaption{}
    \label{dataratedynamicdifferentlambda}
\end{subfigure}
\caption{The effect of deactivating the elements of STAR-RIS for $N=144$ in a mobile DU scenario: (a) Data rate plot for $\mu=0.3$ (better visible in colour). Here, $r_j$ denotes the DL data rate of the $j$-th DU and ($\frac{1}{6} \sum_{j=1}^{6} r_j$) represents the average DL data rate across the six DUs. (b) Number of active STAR-RIS elements for different $\mu$ values vs. episode index, and (c) Average data rate vs. episode index for different $\mu$ values. Even in dynamic scenarios, the average data rate remains close for $\mu$ values of 0, 0.1, and 0.3.  The observed fluctuations in the graphs are due to the dynamic nature of the scenario, where the phase-shift and element assignment variables of the STAR-RIS are continuously optimized using DRL.}
\label{fig: mobile}
\end{figure*}

\subsubsection{Proposed DRL-based joint optimization of element assignment variable and STAR-RIS phase shift}
Instead of an equal number of STAR-RIS elements per DU, the proposed DRL method learns to predict $\mathbf{A}$ (Eq. (\ref{eqn:3})) based on the channel condition. We consider $N=144$ ($N^{t}=N^{r}=72$) STAR-RIS elements. The reward function given by Eq. (\ref{eq:14}) is equivalent to Eq. (\ref{eq:rew}) for $\mu=0$, and with this reward function, none of the STAR-RIS elements are deactivated. To ensure that the constraints \ref{eq:optimization:2a} and \ref{eq:optimization:3a} hold, i.e., each DU is served by at least one element, we fix at least one element for each DU, and the assignment of the rest of the elements is predicted by the DRL algorithm.
In Fig. \ref{fig12by12-learn}, it can be observed that predicting the element assignment matrix using the proposed DRL method improves average data rates, mainly benefiting DU $3$ and DU $6$, spatially distant from STAR-RIS, with negligible degradation in the data rates of the other DUs. The method proposed (Fig. \ref{fig12by12-learn}) can be compared to the equal partitioning given in Fig. \ref{fig12by12}. Figs. \ref{fig12by12} and \ref{fig12by12-learn} are generated under identical conditions, including the same DU positions, channel realizations, $N=144$ STAR-RIS elements, and DDPG training parameters (Table \ref{tab:allmethods1}), with $\mu=0$ so that element deactivation is not considered. Observe that the use of more STAR-RIS elements for DUs is not always beneficial. As the method intelligently allocates more elements to distant DUs having a poorer channel condition, DUs with comparatively better channel conditions are left with a smaller number of elements. The data rates of these DUs (DU $1$, DU $2$, DU $4$, and DU $5$) improve even for a smaller number of STAR-RIS elements. The average data rate increases by $34\%$ compared to the equal partitioning.  So, given that we have more than enough STAR-RIS elements than what is needed to serve all the DUs, it will be of great interest to see if we can selectively deactivate some of the STAR-RIS elements to achieve similar performance to make the system more resource-efficient.

\subsubsection{Selectively deactivating STAR-RIS elements}
We also investigate the effect of deactivating some of the STAR-RIS elements, where we have $N=144$ ($N^{t}=N^{r}=72$) STAR-RIS elements. Using a value of $\mu=0.4$ in Eq. (\ref{eq:14}), we encourage the deactivation of excess STAR-RIS elements and plot the data rates of all DUs and the average data rate in Fig. \ref{IRS12BY12-SHUTDOWN}. 
We obtain a similar average data rate for both $\mu=0$ and $\mu=0.4$, with only about a $3.2\%$ reduction in the average data rate for $\mu=0.4$, as shown in Figs. \ref{fig12by12-learn}  and \ref{IRS12BY12-SHUTDOWN}, respectively. We are also interested in looking at the number of active STAR-RIS elements for different values of $\mu$. Fig. \ref{comparelambda} shows the total number of active STAR-RIS elements vs. the episode index for different $\mu$.  A higher value of $\mu$ encourages the DRL agent to selectively deactivate more STAR-RIS elements. For $\mu=0$, all the elements of STAR-RIS are active. For values of $\mu$ ranging from $0$ to $0.6$, the number of active elements decreases from $144$ to $86$, indicating that the DRL model effectively deactivates elements. With $\mu=0.4$, we can deactivate, on average, almost $34\%$ of the STAR-RIS elements (active elements reduced from 144 to 95), thus improving resource utilization. Fig. \ref{dataratefordifferentlambda} shows the average data rate vs. the episode index for $\mu=0, 0.1, 0.4$ and $\mu=0.6$. It can be observed that the average data rates obtained for $\mu=0$ and $\mu=0.4$ remain relatively close. With $\mu=0.4$, approximately $34\%$ of the STAR-RIS elements can be deactivated, while the average data rate decreases by only about $3\%$, thereby improving resource utilization with only a marginal performance loss. However, when $\mu$ is further increased to $0.6$, although the number of deactivated STAR-RIS elements increases, the average data rate decreases by approximately $7.1\%$. Thus, increasing $\mu$ promotes the deactivation of more STAR-RIS elements, but an excessively large value of $\mu$ results in a reduction in the data rate, indicating a trade-off between resource utilization and system performance.
\subsection{Study the impact of the element assignment variable on achieving fair and high data rates among mobile DUs}
We now evaluate the effectiveness of the proposed method for mobile DUs, using the random waypoint (RWP) mobility model. This model is widely appealing due to its simplicity, ease of implementation, and ability to represent unrestricted mobility patterns \cite{johnson1996dynamic}. In this model, the movement of the entity is characterized by random changes in position and orientation. The DUs move in a total square area of $100$m$^{2}$ with an average displacement of $1$m per time step, and at the beginning of each episode, the DU positions are randomly initialized. At each time step, the displacement is drawn as  $d_r \sim \mathrm{Rayleigh}\!\left(1/\sqrt{2\pi\lambda_1}\right)$, with $\lambda_1=0.25$, corresponding to an average displacement of $1$m per time step. The motion is restricted to the horizontal plane by setting $d_\vartheta=0$, while the azimuthal direction is drawn as $d_\varphi \sim \mathcal{U}(0,2\pi)$. After each position update, the DU-BS and DU-STAR-RIS distances, path losses, and angles are updated accordingly (refer to the definitions accompanying Eqs (\ref{eqn:10})-(\ref{eqn:11})). The corresponding LoS components in Eqs.(\ref{eqn:10})-(\ref{eqn:11}) are recomputed, while the NLoS components are independently regenerated according to the Rician channel model. Hence, the channel realization at each time step reflects the current DU location. The BS-STAR-RIS channel remains unchanged since both the BS and the STAR-RIS are stationary. Here, we consider the scenario of $N=144$ STAR-RIS elements. The average data rate of the mobile DUs vs. the episode index is plotted in Fig. \ref{IRS12BY12-mobile}. It can be seen that, even though our channels are dynamically fluctuating, we still achieve fair and high data rates for all DUs. Fig. \ref{comparelambdadynamic} illustrates the total number of active STAR-RIS elements vs. the episode index for different values of $\mu$. In this, we examine the cases of $\mu=0, 0.1, 0.3$ and $\mu=0.6$. Due to user mobility, the number of elements allocated to DUs changes over time, resulting in a fluctuating plot vs. the episode index. With $\mu=0.1$, an average of $126$ elements actively serve the DUs. When $\mu=0.3$, the average number of activated elements decreases to $110$, corresponding to the deactivation of approximately $23\%$ of the STAR-RIS elements. Furthermore, for $\mu=0.6$, the average number of activated elements decreases to $100$, corresponding to approximately $31\%$ element deactivation. Fig. \ref{dataratedynamicdifferentlambda} shows the average data rate vs. the training episode index for $\mu=0, 0.1, 0.3$, and $0.6$. It can be observed that the average data rates for $\mu=0$ and $\mu=0.3$ remain relatively close, with only about a $3.5\%$ reduction in average data rate for $\mu=0.3$, indicating that the DRL agent can deactivate excess STAR-RIS elements with only a marginal loss in data-rate performance. However, when $\mu$ is further increased to $0.6$, although the percentage of deactivated elements increases to approximately $31\%$, the average data rate decreases by about $12.7\%$. Thus, our proposed model effectively utilizes DRL, ensuring fair and high data rates for all DUs in both the reflection and transmission spaces, while improving the resource efficiency of the STAR-RIS.
\subsection{Fairness Evaluation Using Jain's Fairness Index (JFI)}
\begin{figure}
    \centering
\includegraphics[width=3.6in, height=1.7in]{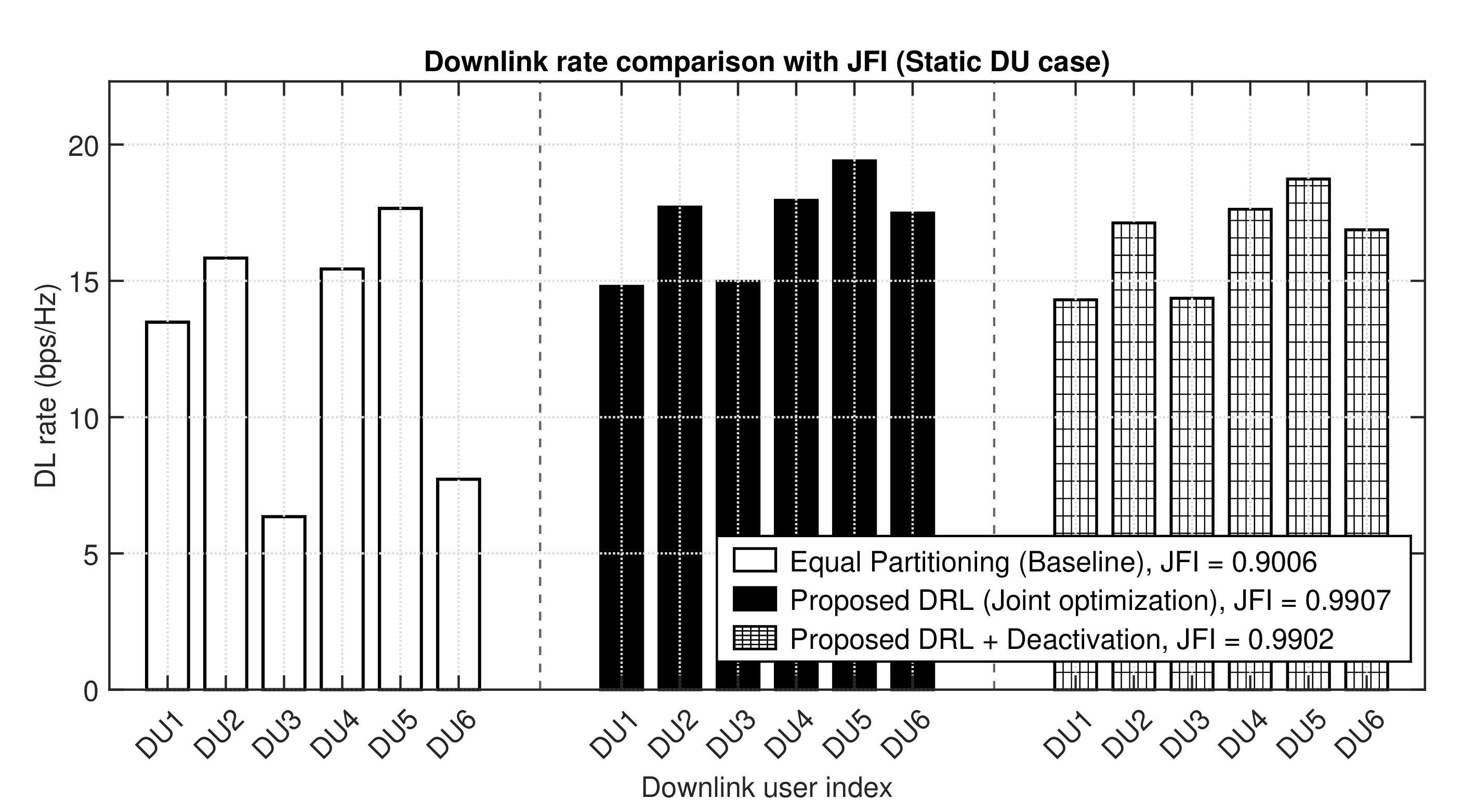}
    \caption{JFI comparison for the three cases under the static DU scenario. The equal-partitioning baseline achieves JFI = 0.9006, the proposed DRL-based joint optimization achieves JFI = 0.9907, and the proposed DRL with selective deactivation achieves JFI = 0.9902.}
    \label{jfistatic}   
\end{figure}

\begin{figure}
    \centering
\includegraphics[width=3.6in, height=1.7in]{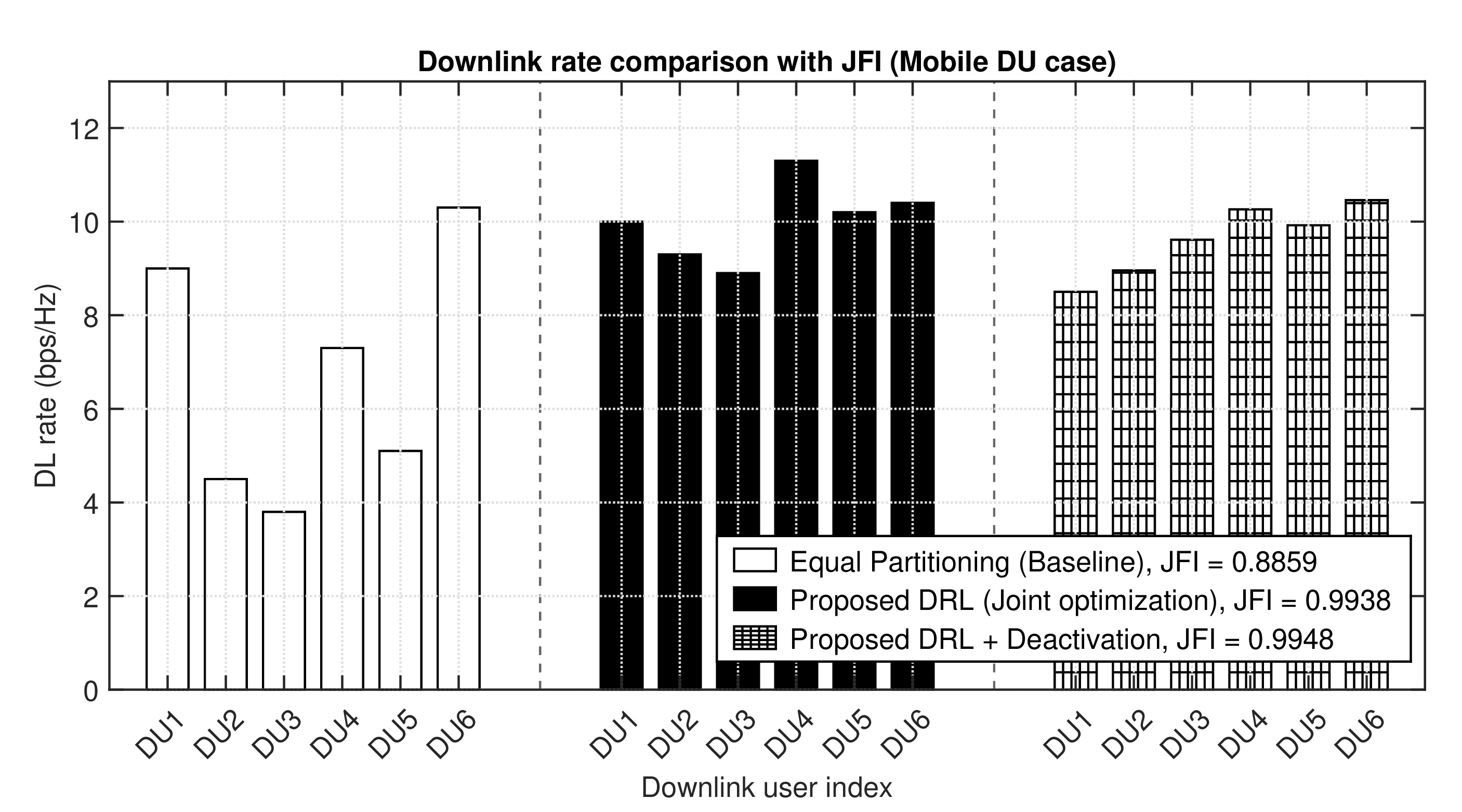}
    \caption{JFI comparison for the three cases under the mobile DU scenario. The
equal-partitioning baseline achieves JFI = 0.8859, the proposed DRL scheme achieves
JFI = 0.9938, and the proposed DRL with selective deactivation achieves JFI = 0.9948.}
    \label{jfimobile}   
\end{figure}

To evaluate the fairness of the DL data rates among DUs, JFI is used \cite{10271700}. Let $J$ denote the total number of DUs. Using the DL data rate values in the reflection and transmission spaces, the JFI is computed as \cite{jain1984quantitative},
\begin{equation}
JFI=\frac{\left(\sum\limits_{k} r^{r}_{k}+\sum\limits_{l} r^{t}_{l}\right)^{2}}
{J\left(\sum\limits_{k}\left(r^{r}_{k}\right)^{2}+\sum\limits_{l}\left(r^{t}_{l}\right)^{2}\right)}, \quad \frac{1}{J}\leq JFI\leq 1,
\end{equation}
where $JFI=1$ indicates perfectly fair rate allocation among all DUs. In this work, JFI is computed using the DL rates of all DUs under the final converged policy. The fairness comparison for different STAR-RIS allocation strategies under static and mobile DU scenarios is given in Fig. \ref{jfistatic} and Fig. \ref{jfimobile}, respectively.
The proposed DRL-based joint optimization of the element assignment variable and STAR-RIS phase shifts improves fairness compared to the baseline of equal partitioning in both static and mobile DU scenarios. In the equal partitioning case, the DRL algorithm optimizes only the STAR-RIS phase shifts, resulting in a comparatively uneven rate distribution among DUs. Moreover, fairness remains consistently high even with selective element deactivation in both static and mobile scenarios, confirming that fairness is preserved while improving resource utilization. It should be noted that for the mobile case (Fig. \ref{jfimobile}), since the DU positions are randomly initialized at the beginning of each episode and subsequently updated according to the RWP mobility model, modest JFI variations are indeed expected.

\begin{figure}
    \centering
\includegraphics[width=3.5in, height=1.8in]{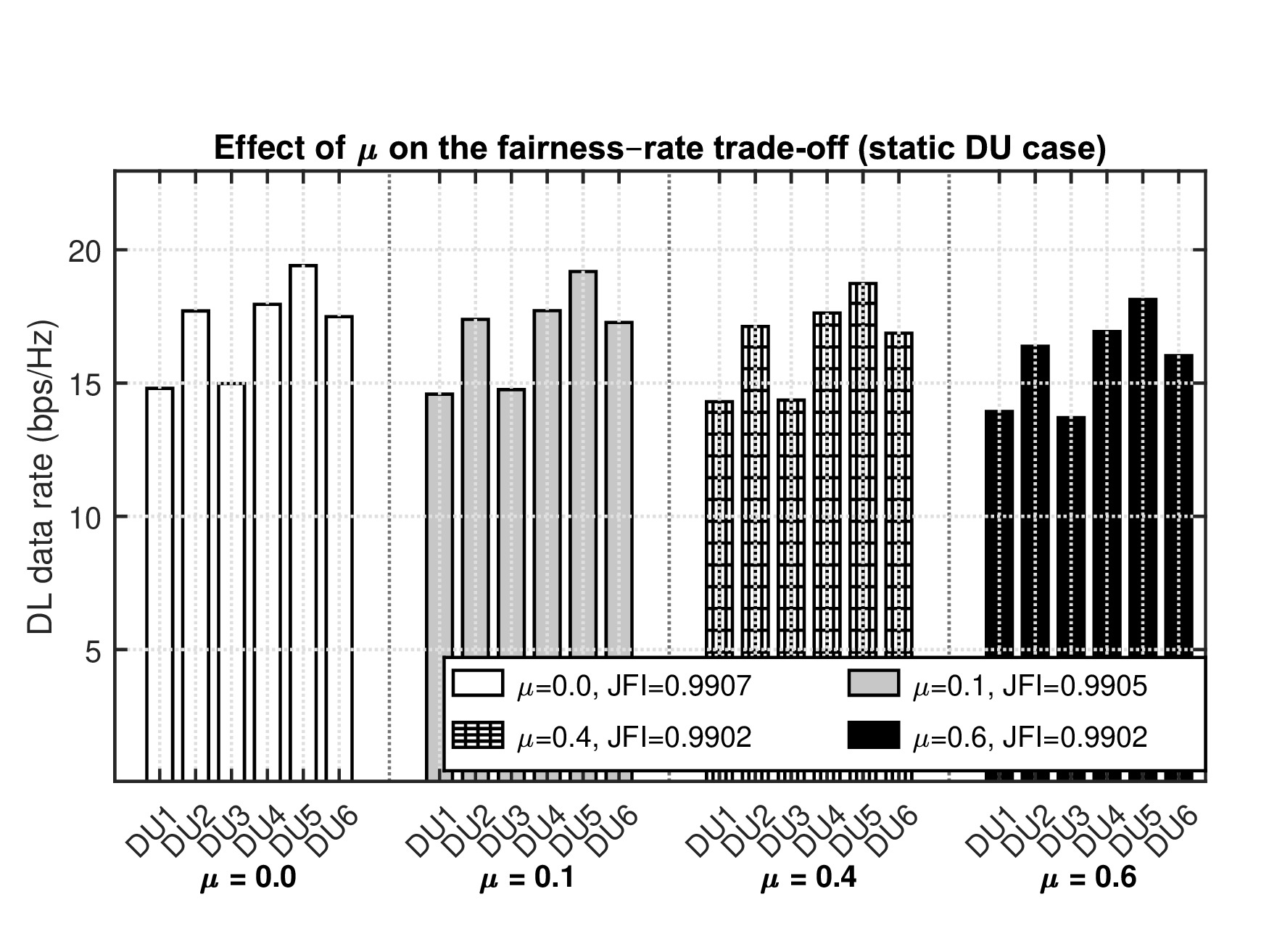}
    \caption{Effect of the weighting factor $\mu$ on the DL data rate and JFI for the static DU scenario. The JFI values for $\mu=0$, $0.1$, $0.4$, and $0.6$ are $0.9907$, $0.9905$, $0.9902$, and $0.9902$, respectively.}
    \label{jfistaticmu}   
\end{figure}
\begin{figure}
    \centering
\includegraphics[width=3.5in, height=1.8in]{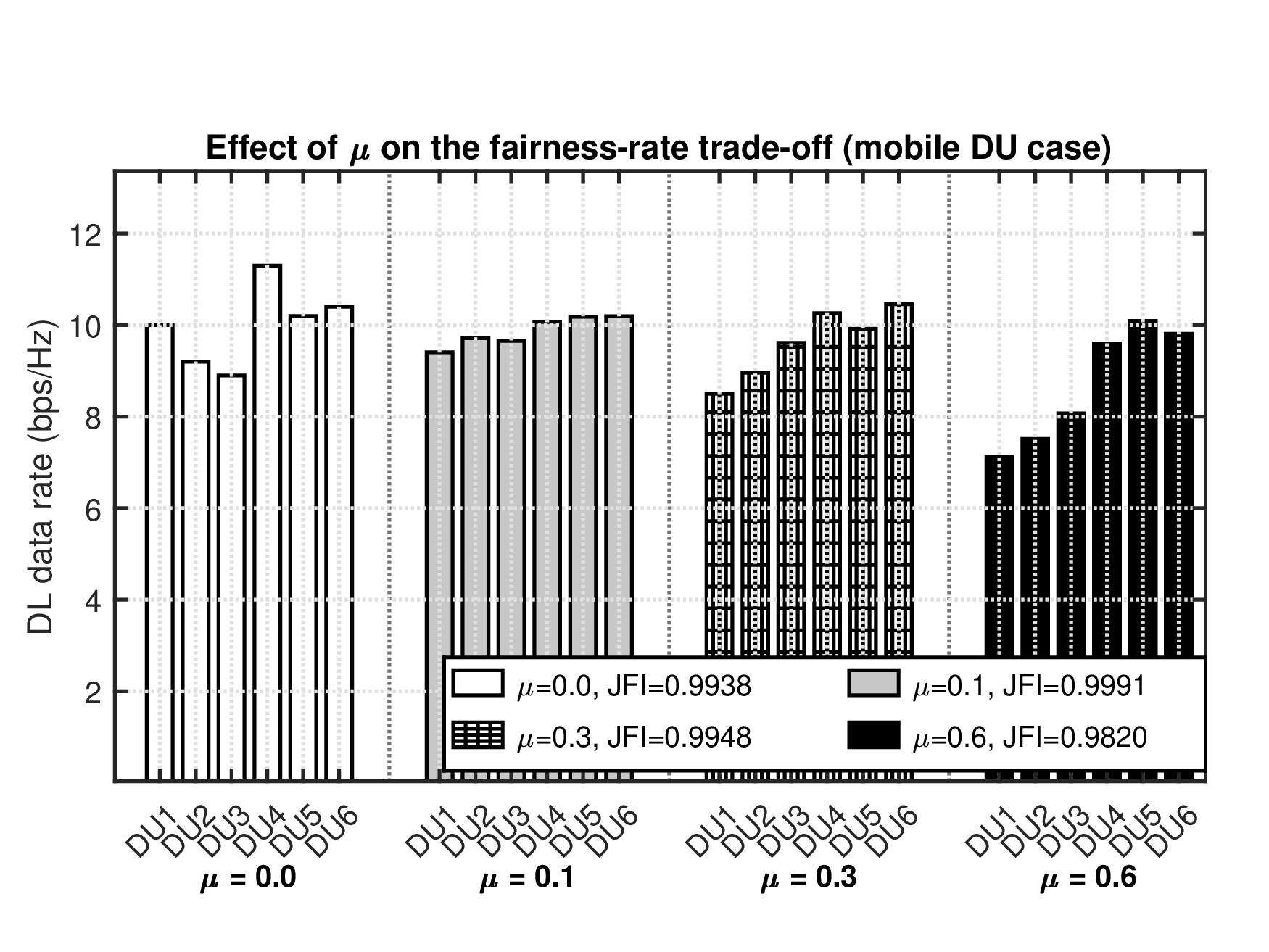}
    \caption{Effect of the weighting factor $\mu$ on the DL data rate and JFI for the mobile DU scenario. The JFI values for $\mu=0$, $0.1$, $0.3$, and $0.6$ are $0.9938$, $0.9991$, $0.9948$, and $0.9820$, respectively.}
    \label{jfimobilemu}   
\end{figure}
Fig.~\ref{jfistaticmu} and Fig.~\ref{jfimobilemu} show the effect of the weighting factor $\mu$ on the DL data rate and JFI for the static and mobile DU scenarios, respectively. For the static case, the average DL data rate decreases from $17.056$ to $16.817$, $16.504$, and $15.850$ bps/Hz as $\mu$ increases from $0$ to $0.1$, $0.4$, and $0.6$, while the corresponding JFI remains nearly unchanged at $0.9907$, $0.9905$, $0.9902$, and $0.9902$. For the mobile case, the average DL data rate decreases from $9.970$ to $9.870$, $9.619$, and $8.700$ bps/Hz as $\mu$ increases from $0$ to $0.1$, $0.3$, and $0.6$, with the corresponding JFI changes from $0.9938$ to $0.9991$, $0.9948$, and $0.9820$. These results show that moderate values of $\mu$ enable greater element deactivation with only a small reduction in data rate, while fairness among the DUs remains consistently high across the considered range of $\mu$.

\subsection{Computational Requirements of the Proposed Algorithm and FLOPs Comparison}
In these simulations, we have considered three DUs in the reflection space and three DUs in the transmission space. Recall that (cf. Section~\ref{sec:Deep Reinforcement Learning based solution}) to address the huge action space involving the STAR-RIS phase shift and element assignment variable, we consider a two-part neural network, having a feed-forward FE and two SNs. At time $t'$, the FE processes the state $\mathbf{u}_{0}$=$\mathbf{s}^{\{ t'\}}$, with dimension $6+2N$. The first six entries corresponds to the observed SINR values (resulting SINR in the DUs in both the transmission ($\gamma_{t}^{\{t'-1\}}$) and the reflection spaces ($\gamma_{r}^{\{t'-1\}}$)) at time ($t'-1$). The next $2N$ entries correspond to the phase shift and the element assignment variable of STAR-RIS, respectively. Assume that each two-layer FE layer has $T$ neurons. The total number of floating-point operation (FLOP) performed at the FE is $2[(6 +2N)T+T^{2}]$. The FLOP for the first and second SNs is $2TN$. A practical computational complexity analysis of the proposed methods, through a run-time based comparison is given in Sec. \ref{sec:Comparison of the proposed DRL method with the hybrid approach}.

\subsection{Effect of Imperfect CSI on Performance}
To characterize the performance of the proposed method under imperfect CSI, we carry out the following experiment. For a channel $\mathbf{h}
\in \{\mathbf{g}_{k}^{r},\mathbf{h}_{k}^{r}, \mathbf{g}_{l}^{t},\mathbf{h}_{l}^{t}, \mathbf{G}\}$, the channel estimation error (CEE) may be expressed as, 
\begin{equation}
\mathrm{CEE} = 10\log_{10}\!\left(
\frac{\mathbb{E}\!\left[\|\mathbf{h}-\hat{\mathbf{h}}\|_2^2\right]}
{\mathbb{E}\!\left[\|\mathbf{h}\|_2^2\right]}
\right).
\label{eq:cee}
\end{equation}
The BS-STAR-RIS, BS-DU, and STAR-RIS-DU channels in the reflection space and transmission space are modeled using an additive CEE model, where 
\begin{equation}
\hat{\mathbf{h}} = \mathbf{h} + \Delta\mathbf{h}, \qquad 
\Delta\mathbf{h} \sim \mathcal{CN}(\mathbf{0},\,\sigma_e^2 \mathbf{I}).
\end{equation}
Here, $\hat{(\cdot)}$ denotes the estimated channel, $(\cdot)$ is the true channel, $\Delta(\cdot)$ represents the estimation error \cite{10623434} and $\sigma_e^2$ denotes the estimation error variance. Fig. \ref{fig: LEARNABLE1} shows the average DL data rate vs. the episode index for imperfect CSI with CEE = -$10$ dB for the static DU scenario. A CEE value of -$10$ dB indicates that the average power of the CEE is $10$\% of the true channel power. It can be seen that the proposed DRL-based method (by joint optimization of element assignment variable and STAR-RIS phase shift) successfully converges even in the presence of CEE. Although the data rate associated with imperfect CSI in Fig. \ref{fig: LEARNABLE1} is slightly lower compared to the one with perfect CSI in Fig. \ref{fig12by12-learn}, the learning process remains stable, and the convergence trend is clearly preserved. Fig. \ref{fig: MOBILE1} corresponds to the mobile DU scenario under the same imperfect CSI condition (CEE = -$10$ dB). A similar convergence behavior is observed, indicating that the learning agent effectively adapts to both user mobility and imperfect CSI.

\begin{figure}
\centering
    \pgfplotstableread[col sep = comma]{./data/IRS12BY12LEARNABLEA1.csv}\datatableone
    \tikzstyle{mark_style} = [mark size={3.0}, mark repeat=20, mark phase=1]              
    \begin{tikzpicture}[thick,scale=1.7]
    \begin{axis}[
        width=4.6cm,
        height=4.4cm,
        xmin=0,
        xmax=210,
        ymin=0,
        ymax=18.5,
        grid=major,
        xlabel={Episode index},
        ylabel={DL rate (bps/Hz)},
       xlabel style={at={(0.50,0.03)}, yshift=5pt},
        ylabel style={at={(0.25,0.50)}, yshift=-3pt},
        label style={font=\tiny, scale=0.8},
        tick label style={font=\tiny, scale=0.7},
        legend pos=south west,
        legend cell align={left},
        legend style={at={(0.31,0)}, fill opacity=0.45, draw opacity=1.0, text opacity=1.0, font=\tiny, scale=0.8, nodes={scale=0.8}}
        ]
       
         \addplot[black!50, solid, thin, mark=oplus*, mark size={2.0},every mark/.append style={solid, fill=cyan!100, opacity=1}, mark repeat=50, opacity=0.9
                ] 
            table [x=x_data, y expr=\thisrowno{1}/1000, col sep=comma]{\datatableone};
        \addlegendentry{$r_{1}$};

        \addplot[black!50, solid, thin, 
                mark=pentagon*, mark size={2.2}, every mark/.append style={solid, fill=blue!100, opacity=1}, mark repeat=43,  opacity=0.9
                ] 
            table [x=x_data, y expr=\thisrowno{2}/1000, col sep=comma]{\datatableone};
        \addlegendentry{$r_{2}$};

        \addplot[black!50, solid, thin, 
                mark=triangle*, mark size={2.0},every mark/.append style={solid, fill=violet!100, opacity=1}, mark repeat=40,  opacity=0.9
                ] 
            table [x=x_data, y expr=\thisrowno{3}/1000, col sep=comma]{\datatableone};
        \addlegendentry{$r_{3}$};

        \addplot[black!50, solid, thin, 
                mark=diamond*, mark size={2.0}, every mark/.append style={solid, fill=orange!100, opacity=1}, mark repeat=48,  opacity=0.9
                ] 
            table [x=x_data, y expr=\thisrowno{4}/1000, col sep=comma]{\datatableone};
        \addlegendentry{$r_{4}$};

        \addplot[black!50, solid, thin, 
                mark=o, mark size={2.0}, every mark/.append style={solid, fill=white!100, opacity=1}, mark repeat=30,  opacity=0.9
                ] 
            table [x=x_data, y expr=\thisrowno{5}/1000, col sep=comma]{\datatableone};
        \addlegendentry{$r_{5}$};

        \addplot[black!50, solid, thin, 
                mark=triangle*, mark size={2.0}, every mark/.append style={solid, fill=red!100, rotate=180, opacity=1}, mark repeat=37,  opacity=0.9
                ] 
            table [x=x_data, y expr=\thisrowno{6}/1000, col sep=comma]{\datatableone};
        \addlegendentry{$r_{6}$};

        \addplot[black!50, solid, thin, 
                mark=square*, mark size={2.0},every mark/.append style={solid, fill=green!100, opacity=1}, mark repeat=50,  opacity=0.9
                ] 
            table [x=x_data, y expr=\thisrowno{7}/1000, col sep=comma]{\datatableone};
        \addlegendentry{$\frac{1}{6} \sum r_j$};

    \end{axis}
\end{tikzpicture}
    \caption{Average DL data rate vs. episode index under imperfect CSI with CEE = -$10$ dB for static DU scenario ($N = 144$ STAR-RIS elements). Here, $r_j$ denotes the DL data rate of the $j$-th DU and ($\frac{1}{6} \sum_{j=1}^{6} r_j$) represents the average DL data rate across the six DUs.} 
\label{fig: LEARNABLE1}
\end{figure}

\begin{figure}
\centering
    \pgfplotstableread[col sep = comma]{./data/IRS12BY12MOBILE1.csv}\datatableone
    \tikzstyle{mark_style} = [mark size={3.0}, mark repeat=20, mark phase=1]              
    \begin{tikzpicture}[thick,scale=1.7]
    \begin{axis}[
        width=4.6cm,
        height=4.4cm,
        xmin=0,
        xmax=210,
        ymin=0,
        ymax=9.5,
        grid=major,
        xlabel={Episode index},
        ylabel={DL rate (bps/Hz))},
        xlabel style={at={(0.50,0.03)}, yshift=5pt},
        ylabel style={at={(0.25,0.50)}, yshift=-5pt},
        label style={font=\tiny, scale=0.8},
        tick label style={font=\tiny, scale=0.7},
        legend pos=south west,
        legend cell align={left},
        legend style={at={(0.35,0)}, fill opacity=0.45, draw opacity=1.0, text opacity=1.0, font=\tiny, scale=0.75, nodes={scale=0.75}}
        ]
         \addplot[black!50, solid, thin, mark=oplus*, mark size={2.0},every mark/.append style={solid, fill=cyan!100, opacity=1}, mark repeat=51, opacity=0.9
                ] 
            table [x=x_data, y expr=\thisrowno{1}/1000, col sep=comma]{\datatableone};
        \addlegendentry{$r_{1}$};

        \addplot[black!50, solid, thin, 
                mark=pentagon*, mark size={2.2}, every mark/.append style={solid, fill=blue!100, opacity=1}, mark repeat=43,  opacity=0.9
                ] 
            table [x=x_data, y expr=\thisrowno{2}/1000, col sep=comma]{\datatableone};
        \addlegendentry{$r_{2}$};

        \addplot[black!50, solid, thin, 
                mark=triangle*, mark size={2.0},every mark/.append style={solid, fill=violet!100, opacity=1}, mark repeat=40,  opacity=0.9
                ] 
            table [x=x_data, y expr=\thisrowno{3}/1000, col sep=comma]{\datatableone};
        \addlegendentry{$r_{3}$};

        \addplot[black!50, solid, thin, 
                mark=diamond*, mark size={2.0}, every mark/.append style={solid, fill=orange!100, opacity=1}, mark repeat=48,  opacity=0.9
                ] 
            table [x=x_data, y expr=\thisrowno{4}/1000, col sep=comma]{\datatableone};
        \addlegendentry{$r_{4}$};

        \addplot[black!50, solid, thin, 
                mark=o, mark size={2.0}, every mark/.append style={solid, fill=white!100, opacity=1}, mark repeat=30,  opacity=0.9
                ] 
            table [x=x_data, y expr=\thisrowno{5}/1000, col sep=comma]{\datatableone};
        \addlegendentry{$r_{5}$};

        \addplot[black!50, solid, thin, 
                mark=triangle*, mark size={2.0}, every mark/.append style={solid, fill=red!100, rotate=180, opacity=1}, mark repeat=37,  opacity=0.9
                ] 
            table [x=x_data, y expr=\thisrowno{6}/1000, col sep=comma]{\datatableone};
        \addlegendentry{$r_{6}$};

        \addplot[black!50, solid, thin, 
                mark=square*, mark size={2.0},every mark/.append style={solid, fill=green!100, opacity=1}, mark repeat=50,  opacity=0.9
                ] 
            table [x=x_data, y expr=\thisrowno{7}/1000, col sep=comma]{\datatableone};
        \addlegendentry{$\frac{1}{6} \sum r_j$};
    \end{axis}
\end{tikzpicture}
    \caption{Average DL data rate vs. episode index under imperfect CSI with CEE = -$10$ dB for mobile DU scenario ($N = 144$ STAR-RIS elements). Here, $r_j$ denotes the DL data rate of the $j$-th DU and ($\frac{1}{6} \sum_{j=1}^{6} r_j$) represents the average DL data rate across the six DUs.} 
\label{fig: MOBILE1}
\end{figure}

\subsection{Comparison of the proposed methods}
We compare the performance and the computational characteristics of our proposed methods for a stationary multi-user scenario. For this, we have taken $N=36$ STAR-RIS elements. We consider three DUs in the reflection space and three DUs in the transmission space. First, we compare the performance of the DRL method to that of Dinkelbach's algorithm to solve the STAR-RIS phase shifts alone. 
\begin{itemize}
    \item DRL-ZF-PS: Phase shifts optimized using DRL with ZF beamforming;
    \item DRL-MRT-PS: Phase shifts optimized using DRL with MRT beamforming;
    \item Dinkelbach-MRT-PS: Phase shifts optimized using Dinkelbach's algorithm with MRT;
\end{itemize}
Next, we have compared the proposed DRL method to the proposed hybrid DRL approach.
\begin{itemize}
    \item DRL-ZF: One of the proposed DRL-based methods that jointly optimizes STAR-RIS phase shifts and element assignment with ZF beamforming;
    \item DRL-MRT: One of the proposed DRL-based methods that jointly optimizes STAR-RIS phase shifts and element assignment with MRT beamforming;
    \item Hybrid-DRL-MRT: One of the proposed hybrid approaches, where Dinkelbach's algorithm optimizes phase shifts, and DRL iteratively finds the element assignment variable with MRT beamforming;
\end{itemize}

\subsubsection{Comparison of the DRL with Dinkelbach's for phase shift optimization}
In this section, we evaluate the performance of the DRL method, using either ZF or MRT beamforming separately and then comparing them with a Dinkelbach-based iterative optimization algorithm, which uses MRT beamforming. For this method, we assume that the element assignment variables are set to $1$ for a fixed set of $6$ STAR-RIS elements per DU, resulting in an equal partitioning of all $N=36$ elements among the $6$ DUs. Fig. \ref{fig: ALL} shows the average DL data rates for the training episode index. We observe that the DRL-ZF-PS achieves a final average data rate of $3.8$ bps/Hz, while its MRT counterpart, namely DRL-MRT-PS reaches $2.8$ bps/Hz. The Dinkelbach-MRT-PS technique rapidly converges within the first few episodes to $2.5$ bps/Hz. DRL-ZF-PS achieves a $52\%$ higher average data rate than Dinkelbach-MRT-PS. However, the main disadvantage of Dinkelbach's scheme is that this method is computationally intensive.

\subsubsection{Comparison of the proposed DRL method to the hybrid approach}
\label{sec:Comparison of the proposed DRL method with the hybrid approach}
The proposed hybrid-DRL-MRT approach is seen to converge quickly, reaching an average data rate of $3.0$ bps/Hz. To give an idea about the computational complexity involved, we give Table \ref{tab:complexity_comparison}, which shows the run-time during the inference, while both are run on the same device (for $1$ episode). The Hybrid-DRL-MRT scheme requires approximately $21$s per episode, corresponding to an average of $21$ms per resource-allocation decision. This computational overhead arises mainly from the iterative convex optimization (implemented using CVXPY) performed at every decision step and increases further with the number of STAR-RIS elements. Also, the hybrid scheme may not be suitable for high-mobility scenarios where the channel coherence time is shorter than its decision latency. It represents an intermediate solution between fully conventional and fully learning-based approaches and is applicable when the STAR-RIS configuration can be updated over a sufficiently slow time scale. The proposed DRL-MRT scheme requires approximately $3.2$ms per decision and eliminates the need for iterative convex optimization during inference, making it more appropriate for mobility and rapidly varying channel conditions. Even though the proposed DRL method achieves a higher average DL data rate, the proposed hybrid DRL converges quickly and can be a practical choice in scenarios, where sufficient computing power is available and slightly eroded performance is acceptable. The proposed DRL-ZF achieves the highest overall performance among all the schemes, reaching approximately $4.4$ bps/Hz, and the conceived DRL-MRT achieves approximately $3.3$ bps/Hz after $40$ episodes, as shown in Fig. \ref{fig: ALL}. The DRL-ZF method has a performance $46\%$ better than hybrid-DRL-MRT. The DRL-ZF also eliminates the need for convex optimization at each decision step, resulting in faster inference than the hybrid DRL scheme. Moreover, the proposed DRL method adapts better to dynamic scenarios, making it suitable for environments with user mobility and rapidly changing channel conditions.

\begin{figure}
\centering
    \pgfplotstableread[col sep = comma]{./data/IRS6BY6ALL.csv}\datatableone
    \tikzstyle{mark_style} = [mark size={3.0}, mark repeat=20, mark phase=1]              
    \begin{tikzpicture}[thick,scale=1.7]
    \begin{axis}[
        width=4.6cm,
        height=4.6cm,
        xmin=0,
        xmax=210,
        ymin=0,
        ymax=5,
        grid=major,
        xlabel={Episode index},
        ylabel={Average DL rate (bps/Hz)},
       xlabel style={at={(0.50,0.03)}, yshift=2pt},
        ylabel style={at={(0.25,0.50)}, yshift=-2pt},
        label style={font=\tiny, scale=0.8},
        tick label style={font=\tiny, scale=0.7},
        legend pos=south west,
        legend cell align={left},
        legend style={at={(0.31,0)}, fill opacity=0.45, draw opacity=1.0, text opacity=1.0, font=\tiny, scale=0.7, nodes={scale=0.7}}
        ]
         \addplot[black!50, solid, thin, mark=oplus*, mark size={2.0},every mark/.append style={solid, fill=cyan!100, opacity=1}, mark repeat=40, opacity=0.9
                ] 
            table [x=x_data, y expr=\thisrowno{1}/1000, col sep=comma]{\datatableone};
       \addlegendentry{DRL-ZF-PS};

        \addplot[black!50, solid, thin, 
                mark=pentagon*, mark size={2.2}, every mark/.append style={solid, fill=blue!100, opacity=1}, mark repeat=40,  opacity=0.9
                ] 
            table [x=x_data, y expr=\thisrowno{2}/1000, col sep=comma]{\datatableone};
       \addlegendentry{DRL-MRT-PS};

        \addplot[black!50, solid, thin, 
                mark=triangle*, mark size={2.0},every mark/.append style={solid, fill=violet!100, opacity=1}, mark repeat=40,  opacity=0.9
                ] 
            table [x=x_data, y expr=\thisrowno{3}/1000, col sep=comma]{\datatableone};
        \addlegendentry{Dinkelbach-MRT-PS};

        \addplot[black!50, solid, thin, 
                mark=diamond*, mark size={2.0}, every mark/.append style={solid, fill=orange!100, opacity=1}, mark repeat=35,  opacity=0.9
                ] 
            table [x=x_data, y expr=\thisrowno{4}/1000, col sep=comma]{\datatableone};
       \addlegendentry{DRL-ZF};

        \addplot[black!50, solid, thin, 
                mark=o, mark size={2.0}, every mark/.append style={solid, fill=white!100, opacity=1}, mark repeat=30,  opacity=0.9
                ] 
            table [x=x_data, y expr=\thisrowno{5}/1000, col sep=comma]{\datatableone};
       \addlegendentry{DRL-MRT};

        \addplot[black!50, solid, thin, 
                mark=triangle*, mark size={2.0}, every mark/.append style={solid, fill=red!100, rotate=180, opacity=1}, mark repeat=37,  opacity=0.9
                ] 
            table [x=x_data, y expr=\thisrowno{6}/1000, col sep=comma]{\datatableone};
       \addlegendentry{Hybrid DRL-MRT};

    \end{axis}
\end{tikzpicture}
    \caption{Average DL data rate vs. episode index for various learning schemes} 
\label{fig: ALL}
\end{figure}

\begin{table}[t]
\caption{Computational complexity (evaluation time) comparison.}
\label{tab:complexity_comparison}
\centering
\renewcommand{\arraystretch}{1.1}
\setlength{\tabcolsep}{3pt}
\begin{tabular}{|>{\centering\arraybackslash}p{1.35cm}|
                >{\centering\arraybackslash}p{2.4cm}|
                >{\centering\arraybackslash}p{2.1cm}|
                >{\centering\arraybackslash}p{1.7cm}|}
\hline
\textbf{Method} & \textbf{Steps per episode} & \textbf{Per step time (s)} & \textbf{Total time (s)} \\
\hline
DRL-MRT        & 1000 & 0.0032 & 3.2 \\
\hline
Hybrid DRL-MRT & 1000 & 0.0210 & 21  \\
\hline
\end{tabular}
\end{table}

\subsection{Impact of residual scattering from unallocated elements}
To investigate the impact of residual scattering from unallocated STAR-RIS elements, we formulate a residual scattering model by distinguishing the allocated and unallocated elements through the element assignment variables. The reflection and transmission side allocation indicators are defined as $a_n^{r}=\sum_{k=1}^{K}\alpha_{kn}^{r}$, and $a_n^{t}=\sum_{l=1}^{L}\alpha_{ln}^{t}$, respectively. Since each STAR-RIS element can be assigned to at most one DU, we have $a_n^{r}+a_n^{t}\leq 1$. The unallocated-element indicator is therefore defined as $\alpha_n^{0}=1-a_n^{r}-a_n^{t}$. Hence, $\alpha_n^{0}=1$ when element $n$ is unallocated and $\alpha_n^{0}=0$ otherwise. The residual scattering matrix on the side $q\in\{r,t\}$ is defined as $\boldsymbol{\Theta}_{0}^{q}
=
\operatorname{diag}
\left(
\alpha_1^{0}\sqrt{\beta_{1}^{q}}e^{j\tilde{\theta}_{0,1}^{q}},
\ldots,
\alpha_N^{0}\sqrt{\beta_{N}^{q}}e^{j\tilde{\theta}_{0,N}^{q}}
\right)$,
where, $\tilde{\theta}_{0,n}^{q}\sim\mathcal{U}[0,2\pi)$ is the uncontrolled phase. Let $\xi_q\in[0,1]$ denote the residual scattering coefficient, where $\sqrt{\xi_q}$ is the corresponding amplitude coefficient. The signal received at the reflection-side DU is modified as
\begin{equation}
\label{eq:received_residual}
\begin{aligned}
y_k
={}&
\left[
\mathbf{g}_{k}^{r}
\left(
\tilde{\boldsymbol{\Theta}}_{k}^{r}
+
\sqrt{\xi_r}\boldsymbol{\Theta}_{0}^{r}
\right)
\mathbf{G}
+
\mathbf{h}_{k}^{r}
\right]
\mathbf{w}_{k}x_{k}\sqrt{\rho_t}
\\
&+
\left[
\mathbf{g}_{k}^{r}
\left(
\tilde{\boldsymbol{\Theta}}_{k}^{r}
+
\sqrt{\xi_r}\boldsymbol{\Theta}_{0}^{r}
\right)
\mathbf{G}
+
\mathbf{h}_{k}^{r}
\right]
\sum_{\substack{j=1\\j\neq k}}^{J}
\mathbf{w}_{j}x_{j}\sqrt{\rho_t}
+n_k .
\end{aligned}
\end{equation}
The corresponding SINR is
\begin{equation}
\label{eq:sinr_residual}
\gamma_k^{r}(\xi_r)
=
\frac{
\rho_t
\|
\mathbf{g}_{k}^{r}
\left(
\tilde{\boldsymbol{\Theta}}_{k}^{r}
+
\sqrt{\xi_r}\boldsymbol{\Theta}_{0}^{r}
\right)
\mathbf{G}\mathbf{w}_{k}
+
\mathbf{h}_{k}^{r}\mathbf{w}_{k}
\|^{2}
}{
\displaystyle
\sum_{\substack{j=1\\j\neq k}}^{J}
\rho_t
\|
\mathbf{g}_{k}^{r}
\left(
\tilde{\boldsymbol{\Theta}}_{k}^{r}
+
\sqrt{\xi_r}\boldsymbol{\Theta}_{0}^{r}
\right)
\mathbf{G}\mathbf{w}_{j}
+
\mathbf{h}_{k}^{r}\mathbf{w}_{j}
\|^{2}
+
\sigma_k^{2}
}.
\end{equation}
A similar expression applies to the transmission-side DUs by replacing $r$ with $t$. When $\xi_r=\xi_t=0$, the above expressions reduce to the original ideal element deactivation case (refer to Eqs. \ref{eqn:7} and \ref{eqn:8}). \label{subsec:residual_scattering}
\subsubsection{Training stability under residual scattering}
In this analysis, we study how the residual scattering from unallocated elements affects the training stability. We consider the static DU scenario and set the deactivation weighting factor to $\mu=0.4$, as defined in the reward function in Eq. \ref{eq:14}. For the sensitivity analysis, we consider
\begin{equation}
\label{eq:sweep}
\xi_r=\xi_t=\xi_q,
\qquad
\xi_q\in\left\{0,10^{-3},10^{-2},10^{-1},0.3\right\},
\end{equation}
corresponding to the ideal element deactivation case to $-30$ dB, $-20$ dB,  $-10$ dB, and to approximately $-5.23$ dB, respectively. Here, the ideal element deactivation case ($\xi_q=0$) refers to the baseline assumption that the unallocated STAR-RIS elements are completely non-contributing and introduce no residual scattering to the received signal. For each residual scattering level $\xi_q$ in \eqref{eq:sweep}, we train a separate DRL policy under the corresponding leakage condition using identical training parameters and channel-generation settings, allowing the phase-shift and element assignment decisions to adapt to each scenario. Fig.~\ref{fig: LEARNABLE11} shows the resultant average DL data rate versus the training episode index for each $\xi_q$. As shown in Fig. \ref{fig: LEARNABLE11}, the proposed scheme maintains stable performance over a wide range of residual scattering levels.  Compared to the ideal element deactivation case, i.e., $\xi_q = 0$, the average DL data rate decreases by only $1.44\%$, $3.47\%$, and $6.79\%$ for $\xi_q = 10^{-3}$, $10^{-2}$, $10^{-1}$, respectively. The corresponding performance reductions increase to approximately $10.85\%$ for $\xi_q=0.3$. Even at $\xi_q=0.3$ the proposed scheme retains approximately $89.15\%$ of the performance achieved under the ideal element deactivation case. These results indicate that the proposed model remains effective even in the presence of residual scattering.

\begin{figure}
\centering
    \pgfplotstableread[col sep = comma]{./data/UNALLOCATEDELEMENTS.csv}\datatableone
    \tikzstyle{mark_style} = [mark size={3.0}, mark repeat=20, mark phase=1]              
    \begin{tikzpicture}[thick,scale=1.7]
    \begin{axis}[
        width=4.4cm,
        height=4.2cm,
        xmin=0,
        xmax=210,
        ymin=0,
        ymax=18.5,
        grid=major,
        xlabel={Episode index},
        ylabel={Average DL data rate (bps/Hz)},
       xlabel style={at={(0.50,0.03)}, yshift=5pt},
        ylabel style={at={(0.25,0.50)}, yshift=4pt},
        label style={font=\tiny, scale=0.8},
        tick label style={font=\tiny, scale=0.7},
        legend pos=south west,
        legend cell align={left},
        legend style={at={(0.21,0)}, fill opacity=0.45, draw opacity=1.0, text opacity=1.0, font=\tiny, scale=0.8, nodes={scale=0.8}}
        ]
       
         \addplot[black!50, solid, thin, mark=oplus*, mark size={2.0},every mark/.append style={solid, fill=cyan!100, opacity=1}, mark repeat=50, opacity=0.9
                ] 
            table [x=x_data, y expr=\thisrowno{1}/1000, col sep=comma]{\datatableone};
        \addlegendentry{$\xi_q=0$ (Ideal case)};

        \addplot[black!50, solid, thin, 
                mark=pentagon*, mark size={2.2}, every mark/.append style={solid, fill=blue!100, opacity=1}, mark repeat=43,  opacity=0.9
                ] 
            table [x=x_data, y expr=\thisrowno{2}/1000, col sep=comma]{\datatableone};
        \addlegendentry{$\xi_q=10^{-3}$};

        \addplot[black!50, solid, thin, 
                mark=triangle*, mark size={2.0},every mark/.append style={solid, fill=violet!100, opacity=1}, mark repeat=40,  opacity=0.9
                ] 
            table [x=x_data, y expr=\thisrowno{3}/1000, col sep=comma]{\datatableone};
        \addlegendentry{$\xi_q=10^{-2}$};

        \addplot[black!50, solid, thin, 
                mark=diamond*, mark size={2.0}, every mark/.append style={solid, fill=orange!100, opacity=1}, mark repeat=48,  opacity=0.9
                ] 
            table [x=x_data, y expr=\thisrowno{4}/1000, col sep=comma]{\datatableone};
        \addlegendentry{$\xi_q=10^{-1}$};

        \addplot[black!50, solid, thin, 
                mark=o, mark size={2.0}, every mark/.append style={solid, fill=white!100, opacity=1}, mark repeat=30,  opacity=0.9] 
           table [x=x_data, y expr=\thisrowno{5}/1000, col sep=comma]{\datatableone};
        \addlegendentry{$\xi_q=0.3$};

    \end{axis}
\end{tikzpicture}
    \caption{Average DL data rate of six DUs under different $\xi_q$.} 
\label{fig: LEARNABLE11}
\end{figure}

\subsection{Comparison with an alternative categorical element assignment representation using Gumbel-Softmax}
The proposed DRL method uses a bin-based mapping to convert the continuous actor output into a discrete STAR-RIS element assignment. To assess whether the performance depends on this specific continuous-to-discrete mapping, we also consider Gumbel-Softmax as an alternative differentiable categorical representation. In general, each STAR-RIS element is represented by $J+1$ categorical outputs, corresponding to the $J$ DUs and one deactivation option, rather than a scalar mapped to bins. During training, Gumbel noise is added to the categorical logits and a soft Gumbel-Softmax representation is generated \cite{jang2016categorical}. This soft representation is provided to the critic, while the corresponding discrete element assignment used by the environment is obtained by selecting the category with the highest categorical value. This provides an explicit categorical assignment rule without requiring bin boundaries or additional boundary-handling rules. For the Gumbel-Softmax implementation, independent standard Gumbel noise $g_{n,j}\sim\mathrm{Gumbel}(0,1)$ is added to the categorical logits during training. The temperature parameter $\tau$ is linearly annealed from $1.0$ to $0.1$ over the first $10^5$ training steps and it is thereafter maintained at $0.1$. Fig. \ref{fig: GUMBEL} shows the comparison between the proposed method and the Gumbel-Softmax-based variant for the static scenario, considering six DUs, element deactivation, and $\mu=0.4$. The proposed method reaches a stable operating region earlier and achieves a slightly higher final average DL data rate of approximately $16.58$ bps/Hz, compared with approximately $16.21$ bps/Hz for the Gumbel-Softmax-based variant, corresponding to an improvement of approximately $2.28\%$. This categorical representation, however, also increases the action dimension. The Gumbel-Softmax-based variant requires a longer training period to stabilize due to its stochastic categorical exploration.
\begin{figure}
\centering
    \pgfplotstableread[col sep = comma]{./data/GUMBEL.csv}\datatableone
    \tikzstyle{mark_style} = [mark size={3.0}, mark repeat=20, mark phase=1]              
    \begin{tikzpicture}[thick,scale=1.7]
    \begin{axis}[
        width=4.4cm,
        height=4cm,
        xmin=0,
        xmax=210,
        ymin=0,
        ymax=19,
        grid=major,
        xlabel={Episode index},
        ylabel={Average DL data rate (bps/Hz)},
        xlabel style={at={(0.50,0.03)}, yshift=5pt},
        ylabel style={at={(0.25,0.50)}, yshift=4pt},
        label style={font=\tiny, scale=0.8},
        tick label style={font=\tiny, scale=0.7},
        legend pos=south west,
        legend cell align={left},
        legend style={at={(0.30,0)}, fill opacity=0.85, draw opacity=1.0, text opacity=1.0, font=\tiny, scale=0.8, nodes={scale=0.8}}
        ]
        \addplot[black!50, solid, thin, 
               mark=oplus*, mark size={2.0},every mark/.append style={solid, fill=red!100, opacity=1.0}, mark repeat=50,  opacity=0.9
                ] 
            table [x=x_data, y expr=\thisrowno{1}/1000, col sep=comma]{\datatableone};
        \addlegendentry{Proposed method};

         \addplot[black!50, solid, thin, 
                mark=square*, mark size={2.0}, every mark/.append style={solid, fill=green!100, opacity=1.0}, mark repeat=60,  opacity=0.9
                ]  
            table [x=x_data, y expr=\thisrowno{2}/1000, col sep=comma]{\datatableone};
        \addlegendentry{Gumbel softmax};


    \end{axis}
\end{tikzpicture}
    \caption{Average DL data rate versus episode index for the deactivation case: proposed method and Gumbel-Softmax comparison.} 
\label{fig: GUMBEL}
\end{figure}

\subsection{Effect of the minimum element assignment constraint on fairness and rate}
To validate the role of the per-DU element assignment constraints in promoting fairness, we generalize constraints \ref{eq:optimization:2a} and \ref{eq:optimization:3a} by replacing the minimum requirement of one element per DU with $N_0$, where $N_0$ denotes the minimum number of elements assigned to each DU. We consider $N_0=0$, which is equivalent to removing the constraints \ref{eq:optimization:2a} and \ref{eq:optimization:3a}, the proposed formulation with $N_0=1$, and $N_0\in\{2,4,8,16\}$. These cases are also compared with equal partitioning, in which the $N=144$ elements are equally divided among the six DUs. Table \ref{tab:constraint_fairness_comparison} presents the average DL rate, JFI, and element allocations obtained for each scheme in the static DU scenario. The case when $N_0=0$ achieves a higher data rate, but the fairness suffers, since it is simply a sum rate optimization. This indicates that without the minimum STAR-RIS element assignment requirement, the DRL agent may favor DUs with better channel conditions to increase the sum rate, resulting in a substantial loss of fairness. The JFI also falls below that of equal partitioning, showing that maximizing the sum data rates without constraints (\ref{eq:optimization:2a}) and (\ref{eq:optimization:3a}) can produce a less balanced rate distribution than equal element partitioning. These results highlight the role of the minimum element assignment constraints in promoting fairness while maintaining a high average DL rate. The proposed objective (with $N_0 = 1$) given in $\mathcal{P}_1^{}$ has a very comparable average data rate with very high JFI. We observe that as the value of $N_0$ increases from $N_0=2$ to $N_0=8$, the JFI keeps improving at the cost of average data rates. When the constraints use $N_0=16$, it moves close to equal partitioning and therefore both the data rates and the JFI reduce. So this observation implies that $N_0$ provides a soft inclusion of min rate into the sum rate expression, and provides a trade-off between sum-rate and min-rate given that the value of $N_0$ is not very high (close to equal partitioning). Thus, $N_0=1$ is enough to bring the min rate into consideration, and then the DRL agent can allocate the rest of the elements successfully such that the distant DUs get more elements than the near DUs and the fairness is maintained.

\begin{table}[t]
\centering
\caption{Effect of $N_0$ on the average DL rate, JFI, and element allocations for the static DU scenario with $N=144$ and $\mu=0$. Results are averaged over four independent runs.}
\label{tab:constraint_fairness_comparison}

\renewcommand{\arraystretch}{1.3}
\setlength{\tabcolsep}{1.8pt}
\footnotesize

\begin{tabular}{|c|c|c|c|c|c|c|c|c|}
\hline
\textbf{Scheme} &
\makecell{\textbf{Avg. DL}\\\textbf{rate}} &
\textbf{JFI} &
\multicolumn{6}{c|}{\textbf{Element allocation}} \\
\cline{4-9}
&
\textbf{(bps/Hz)} &
&
\textbf{DU1} & \textbf{DU2} & \textbf{DU3} &
\textbf{DU4} & \textbf{DU5} & \textbf{DU6} \\
\hline

\makecell{Proposed\\$N_0=1$}
& 17.05 & 0.9907 & 18 & 19 & 35 & 19 & 18 & 35 \\
\hline

\makecell{Equal\\partitioning}
& 12.74 & 0.9006 & 24 & 24 & 24 & 24 & 24 & 24 \\
\hline

$N_0=0$
& 17.90 & 0.8591 & 20 & 39 & 13 & 20 & 41 & 11 \\
\hline

$N_0=2$
& 16.88 & 0.9909 & 20 & 19 & 33 & 19 & 20 & 33 \\
\hline

$N_0=4$
& 16.65 & 0.9910 & 19 & 18 & 35 & 19 & 19 & 34 \\
\hline

$N_0=8$
& 16.01 & 0.9921 & 19 & 20 & 33 & 18 & 20 & 34 \\
\hline

$N_0=16$
& 15.19 & 0.9298 & 24 & 21 & 27 & 23 & 20 & 29 \\
\hline

\end{tabular}
\end{table}

\section{Conclusions}
\label{sec:Conclusion}
We have laid the foundation for adaptive STAR-RIS element activation efficiently to ensure fair data rates for both static and mobile DUs. We formulated a novel optimization problem that maximizes the sum of the DU data rates, subject to a minimum per-DU element assignment constraint that embeds an implicit fairness mechanism into the objective. We solved this problem using a DRL algorithm that jointly determines the allocation of STAR-RIS elements to different DUs and phase shifts of STAR-RIS. We first considered the effect of equal and unequal partitioning of the STAR-RIS elements. Our results showed that the proposed DRL method increases the average data rate by $34\%$ compared to equal partitioning for the same total number of STAR-RIS elements in the static DU scenario. The learnable element assignment variable helps to dynamically allocate the STAR-RIS elements even for mobile DUs, to get fair data rates across all the DUs. Furthermore, by incorporating a deactivation incentive into the DRL reward function, excess STAR-RIS elements can be selectively deactivated, leading to efficient resource utilization. Our simulation results indicate that, on average, we can deactivate $34\%$ of STAR-RIS elements in a static DU scenario and $23\%$ in a moving DU scenario with negligible degradation in the average DL data rate. The proposed DRL method has been compared with a Dinkelbach algorithm and with a novel hybrid DRL technique. The simulation results show that the proposed DRL achieves the highest average DL data rate among all.

 	\bibliographystyle{IEEEtran}
   \bibliography{library}
\end{document}